\documentclass[twocolumn]{aastex631}

\usepackage{soul}

\usepackage{amsmath}

\shorttitle{Modeling the stellar mass distribution of FRB hosts}
\shortauthors{Loudas et al.}
\graphicspath{{./}{figures/}}

\begin{document}

\title{Unveiling the origin of fast radio bursts by modeling the stellar mass and star formation distributions 
of their host galaxies}

\correspondingauthor{Nick Loudas}
\email{loudas@princeton.edu}

\author[0000-0001-7599-6664]{Nick Loudas}
\affiliation{Department of Astrophysical Sciences, Peyton Hall, Princeton University, 4 Ivy Lane, Princeton, NJ 08544, USA}

\author[0000-0001-7931-0607]{Dongzi Li}
\affiliation{Department of Astrophysical Sciences, Peyton Hall, Princeton University, 4 Ivy Lane, Princeton, NJ 08544, USA}

\author[0000-0002-0106-7755]{Michael A. Strauss}
\affiliation{Department of Astrophysical Sciences, Peyton Hall, Princeton University, 4 Ivy Lane, Princeton, NJ 08544, USA}

\author[0000-0001-6755-1315]{Joel Leja}
\affiliation{Department of Astronomy \& Astrophysics, The Pennsylvania State University, University Park, PA 16802, USA}
\affiliation{Institute for Computational \& Data Sciences, The Pennsylvania State University, University Park, PA 16802, USA}
\affiliation{Institute for Gravitation \& the Cosmos, The Pennsylvania State University, University Park, PA 16802, USA}



\begin{abstract}

Diverse formation channels have been proposed to explain the emergence
of fast radio bursts (FRBs), yet their origin remains elusive. 
With improved localization precision, roughly 90 FRBs are now associated 
with host galaxies. SED fitting to the host galaxy photometry reveals their stellar masses ($M_\star$) and star formation rates (SFRs), enabling discrimination between various formation channels.
We conduct an extensive comparison of the stellar mass, SFR and z  distributions of 51 FRB hosts and mock-generated galaxy samples to test whether FRBs trace SFR or $M_\star$.
We incorporate a mass-to-light ratio prescription to address optical selection biases. 
In line with \cite{Sharma2024}, we provide evidence in favor of FRB progenitors tracking SF rather than stellar mass. We show that the shape of the assumed $(M_\star/L_r)_{obs}$ distribution
affects the predictions, bringing the low mass end of the stellar mass distribution closer to the data when accounting for the $\mathrm{SFR} - (M_\star/L_r)_{obs}$ correlation. The K-correction effect in the $r-$band is minimal for galaxies at $z \lesssim 0.7$.
Even if FRBs trace SF, up to $\sim6\%$ of a flux-limited FRB host sample can reside below the star-forming main sequence. Finally, we examine a hybrid model in which a fraction of FRBs tracks stellar mass rather than SF. This fraction can be as large as $\sim(40-50)\%$, suggesting that multiple
formation channels are still consistent with observations. The toolkit developed in this work is publicly available (\texttt{GALFRB}\footnote{\url{https://github.com/loudasnick/GALFRB}}  code), offering a straightforward way to generate mock galaxy samples suitable for direct comparisons with future FRB host galaxy data.
\end{abstract}


\section{Introduction} \label{sec1}

Since the discovery of extragalactic, millisecond-duration fast radio bursts (FRBs) in 2007 \citep{Lorimer2007} a number of mechanisms
have been proposed to generate such immense radio wave bursts
(for a recent review see \citealt{Platts2019}; see also \citealt{Zhang2023}). 
Their energetics ($S_v \sim \text{a few dozens Jy}$ at $v \sim 1.4 ~\mathrm{GHz}$), durations ($W \sim \mathrm{ms}$), high linear and/or circular polarization fractions (e.g., \citealt{pandhi2024, Ng2024}), and large Faraday rotation magnitudes ($\mathrm{RM} \sim 10^{2-4} ~\mathrm{rad ~ m^{-2}} $; \citealt{Michilli2018}) imply strongly magnetized, compact progenitors. In addition, their large dispersion measures ($\mathrm{DM} \sim 10^{2-3} ~\mathrm{cm^{-3}~ pc}$) and high galactic latitudes \citep[e.g.,][]{Champion2016} suggest that they lie at cosmological distances. 

Most detected FRBs are apparent one-off (e.g., cataclysmic) events, though a considerable number of FRBs emit repeated bursts on a range of time scales. The first repeating FRB (rFRB 20121102A) was reported by \cite{Spitler2016}. Its repeating nature enabled precise localization through interferometry, leading to its unambiguous association with a low-metallicity, dwarf star-forming galaxy at $z=0.19$ \citep{Tendulkar2017}, further establishing the extragalactic origin of FRBs.

Since then, considerable progress
has been made towards the identification of their progenitors, yet their origin remains highly elusive.
The association of the Galactic FRB 200428 
with the Galactic magnetar SGR J1935+2154 \citep{Bochenek2020, CHIME2020}, embedded into a supernova remnant (SNR; \citealt{Kothes2018}), along with its temporal coincidence with an X-ray burst (see, e.g., \citealt{Mereghetti2020}), 
has positioned the young strongly magnetized neutron star (NS) as the leading candidate for FRB progenitors.

The proposed sources of FRBs can be broadly categorized into two classes based on the time delay between the progenitor's formation and the observed FRB emission:~(1) those originating from the collapse of massive stars (short-time-scale channel), such as young isolated magnetars in supernova remnants (for a review, see, \citealt{Zhang2023}), and (2) those formed or influenced by dynamical interactions (delayed or dynamical channel). The latter category includes magnetars or pulsars formed or recycled through mergers \citep{Kremer2021,Li2024}, as well as binary compact object systems undergoing mergers or interactions (for a review, see, \citealt{Cordes2019,Zhang2023}). The first class is tied to star formation (SF) in the past $\lesssim 1 ~\text{Myr}$, as massive stars are short-lived. Conversely, the latter is associated with the stellar mass ($M_\star$) and stellar encounter rate, therefore preferentially resides in massive galaxies and globular clusters. Studying the properties of FRB hosts can yield firm evidence for a particular formation channel. 

Localized FRBs for which host galaxies have been securely identified (see, e.g., \citealt{Gordon2023, Law2024}) enable us to study their galaxy demographics and properties through spectral energy distribution (SED) fitting. 
If the stellar mass distribution matches either that of star-forming or massive galaxies, then the occurrence of FRB is tied to the current SF rate (SFR) or its integral, the accumulation of past SF, 
respectively \citep{Law2024}. Consequently, the distribution of stellar masses and/or SFRs of FRB hosts holds significant discriminatory power for currently viable FRB models.

However, no concordant picture of the FRB progenitors has yet emerged from studies of FRB hosts. Most FRBs are found in star-forming environments \citep{Chatterjee2017, Mannings2021, Bhandari2022, Gordon2023, Sharma2024}. Nevertheless, others have been discovered in quiescent galaxies \citep{Sharma2023, Eftekhari2024}, and one lies in a globular cluster \citep{Kirsten2022}, promoting the idea of multiple formation channels.

\cite{Bochenek2021} showed that the properties of FRB hosts are consistent with core-collapse supernovae (CCSNe) hosts, supporting the young magnetar scenario. \cite{Gordon2023} analyzed 23 FRBs with secure host associations and demonstrated that the vast majority lie along the star-forming main sequence (SFMS) of galaxies, supporting progenitor models with short delay channels. Likewise, \cite{Law2024} analyzed a sample of 11 localized FRBs and found that FRBs trace SFR, although they emphasized that their sample also includes two massive, quiescent galaxies. Recently, \cite{Sharma2024} compiled the largest FRB host sample, consisting of 51 objects in total: 25 new secure FRB hosts and 26 presented by \cite{Gordon2023} and \cite{Bhardwaj2024b}. They compared the properties of FRB hosts with various mock-generated galaxy populations and concluded that FRBs are most likely SF tracers, giving additional evidence in favor of the CCSN magnetar formation channel. 

This type of analysis can suffer from selection biases. FRB host associations are based on optical magnitude-limited surveys, thus proper treatment of this selection bias should be conducted when comparing with mock galaxy populations (see, for example \citealt{Seebeck2021, Sharma2024}). Similarly, when attempting to estimate the FRB event rate, inclination-related biases should be considered because propagation-induced transmission effects, such as dispersion and scattering, in the host galaxy could obscure a fraction of FRBs embedded in disks of edge-on observed galaxies \citep{Bhardwaj2024}.

In this paper, we aim to revisit the results of \cite{Sharma2024} and extend their work by applying a
more detailed modeling of the mass-to-light ratio distribution, while taking into account additional effects, such as the so called K-correction, and testing various spatial distributions of mock generated galaxies to quantify their overall impact on the predicted stellar mass distribution. In addition, we estimate the relative percentage of FRBs tracing stellar mass to those tracing SF for given FRB host galaxy dataset. 
The machinery developed in this paper serves as a flexible prescription for comparing mock galaxy populations with future large samples of FRB hosts, with the aim to provide stringent constraints on the origin of FRBs. 

The layout of this paper is as follows. In Sect.~\ref{sec2}, we describe our approach of generating mock galaxy samples that track a specific physical quantity, such as SFR or stellar mass, and discuss the effect of optical selection biases in modeling FRB host galaxies. In Sect.~\ref{sec3}, we quantify the effects of various ingredients of the prescription, such as the weight function, the $\mathrm{SFR}-M_\star -z$ probability distribution, the shape of the mass-to-light ratio distribution, the spatial distribution of mock galaxies, and the K-correction. In Sect.~\ref{sec4}, we present our results and compare the theoretical predictions with FRB host galaxy data. In Sect.~\ref{sec5}, we discuss our findings and summarize the major conclusions of this work. 

In this paper, we fix the cosmological model to be the standard Flat $\Lambda$CDM model
with parameters: $H_0 = 70 ~\mathrm{km ~s^{-1} ~Mpc^{-1}}$, $\Omega_{m,0}=0.3$, and $\Omega_\Lambda = 0.7$.
The overall results, however, are not sensitive 
to small variations in these parameters. The initial mass function (IMF) used by \cite{Sharma2024} (and thus adopted in this paper) to obtain the posterior distribution of the FRB host properties is that of \cite{Kroupa2001}.

\section{Generating mock galaxy samples} \label{sec2}

Several authors (e.g., \citealt{Bochenek2021, Bhandari2022}) have performed a comparison between the stellar mass distribution of FRB host galaxies and that corresponding to galaxy hosts of other transient classes, such as core-collapse supernovae (CCSNe) or ultra-luminous X-ray sources (ULXs). In contrast, \cite{Law2024} compare the FRB host stellar mass distribution with galaxy samples selected based on either $\mathrm{SFR}$ or stellar mass, 
demonstrating the capability of the stellar mass distribution to discern between populations which track different physical quantities. 

In this paper, we model the stellar mass distribution of FRB hosts by constructing
mock galaxy samples that track either $\mathrm{SFR}$ or stellar mass, and compare them with observations (see \citealt{Sharma2024}).
Below, we provide a prescription for generating mock galaxy samples that trace various physical quantities, while also accounting for optical selection effects in galaxy surveys that could bias 
the observations towards identifying a higher relative fraction of more luminous, 
massive host galaxies. \cite{Sharma2024} took this effect into account by adopting a log-normal mass-to-light ratio distribution of all galaxies to convert stellar masses to apparent magnitudes, assuming that galaxies are uniformly distributed in redshift and ignoring the effect of the $K$-correction. In this study, we carry out more careful modeling and highlight the effects of the model choices.

\subsection{Sampling mock host galaxies for various scenarios} \label{sec2.1}

\subsubsection{Stellar mass function} \label{sec2.1.1}

To generate mock galaxy stellar mass samples, we need to quantify the stellar mass function
of galaxies, namely

\begin{equation}
    \Phi(M_\star; z) ~ d\log M_\star = \frac{dn(M_\star; z)}{d\log M_\star} ~d\log M_\star ,
    \label{eq.1}   
\end{equation}
the number density of galaxies in the logarithmic stellar
mass interval of width $d\log M_\star$ centered on $\log M_\star$ at a given redshift $z$.
A well-established parametric stellar mass function in the literature is the so-called Schechter function 
\citep{Schechter1976}, which reads

\begin{equation}
    \phi(M_\star; \phi_c, \alpha, M_c, z) = \ln{(10)} \phi_c \left(\frac{M_\star}{M_c}\right)^{(\alpha + 1)} e^{-M_\star/M_c},
    \label{eq.2}
\end{equation}
where $\phi_c$ is the normalization, $M_c$ is the so-called cutoff mass, 
and $\alpha$ is the power-law index at low masses. 
In general, the parameters $\phi_c$, $M_c$, and $\alpha$ are redshift-dependent due to cosmic evolution. 
\cite{Bochenek2021} used a sample of stellar masses
obtained through SED fitting of PRIMUS \citep{Coil2011, Cool2013} and SDSS galaxy data in the redshift range $0.2-0.3$ by \cite{Moustakas2013} to fit the parameters of the Schechter function and implemented it into a prescription of correcting FRB host galaxy stellar masses for redshift evolution, although they assumed that the Schechter parameters
are redshift-independent over this limited redshift range.

However, recent studies (see, e.g., \citealt{Wright2018}) suggest that a redshift-dependent, two-component Schechter function
\begin{align}
    \Phi(M_\star; z) = & \phi_1(M_\star; \phi_{c,1}, \alpha_1, M_c, z ) \notag \\ 
    &+ \phi_2(M_\star; \phi_{c,2}, \alpha_2, M_c, z )
\end{align}
with redshift-dependent parameters, provides a significantly
better fit to galaxy data, where $\phi_1$ and $\phi_2$ are given by Eq.~\eqref{eq.2}. Note that $M_c$ is fixed to be the same for the two components.
Following \cite{Sharma2024}, in this work, we employ the continuity model presented by \cite{Leja2020}, which involves a
two-component Schechter function while enabling a smooth redshift evolution of the mass function. It has been obtained by using $\sim 10^5$ galaxies from the 3D-HST \citep{Skelton2014} and COSMOS-2015 \citep{Laigle2016} surveys.
The study provides full posteriors for an evolving stellar mass function valid in the redshift range $(0.2, 3)$. 
Each parameter $\rho_i$ of the Schechter function, i.e., $\phi_{c,1}$, $\phi_{c,2}$, $\alpha_1$, $\alpha_2$, and $M_c$, is assumed to follow a quadratic equation of redshift $z$, that is, $\rho_i(z) = a_{0,i} + a_{1,i} z + a_{2,i} z^2$.
\cite{Leja2020} give the fitted values of all relevant coefficients, and present
a guide to generating a mass function in their appendix B. 

\begin{figure*}
    \centering
    \includegraphics[width=0.32\linewidth]{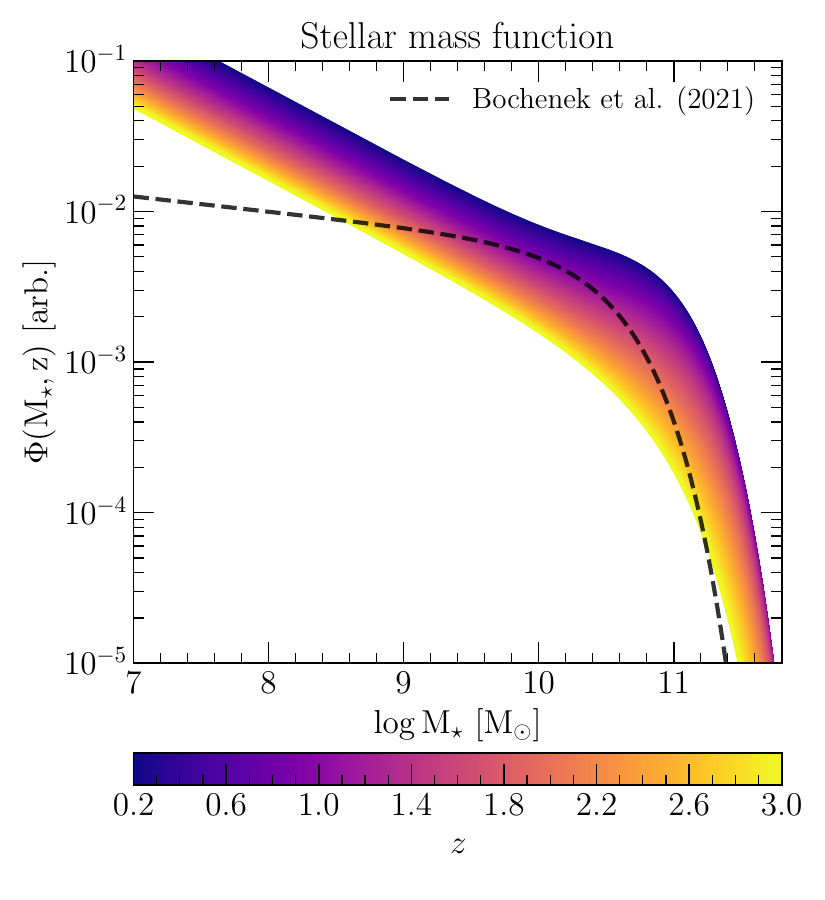}
    \includegraphics[width=0.32\linewidth]{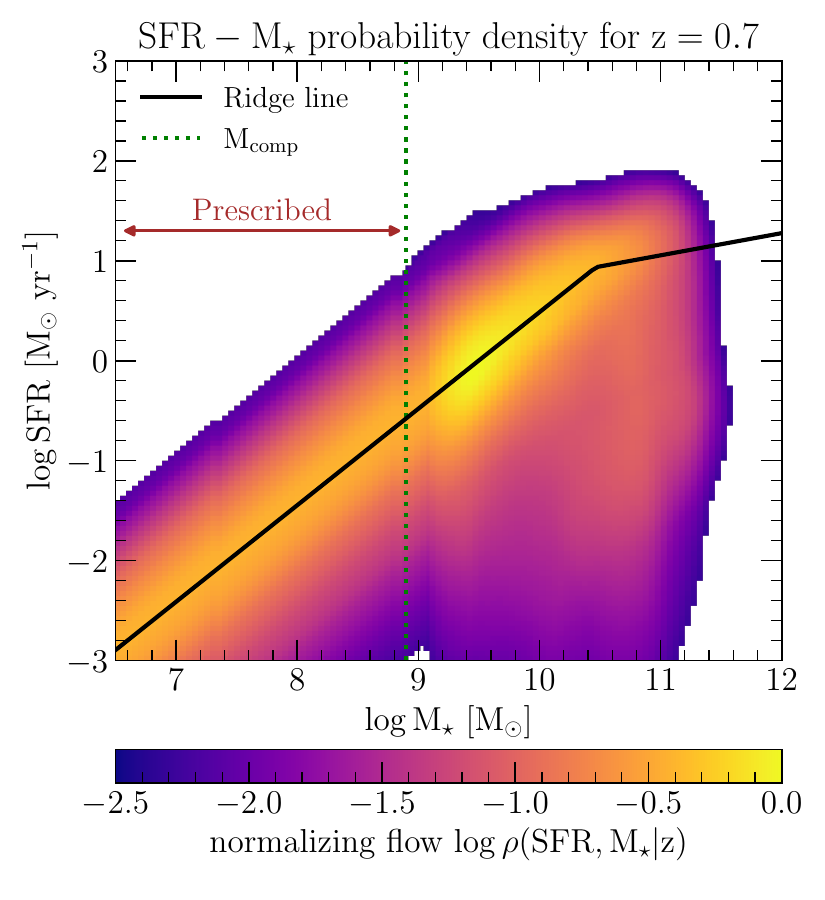}
    \includegraphics[width=0.32\linewidth]{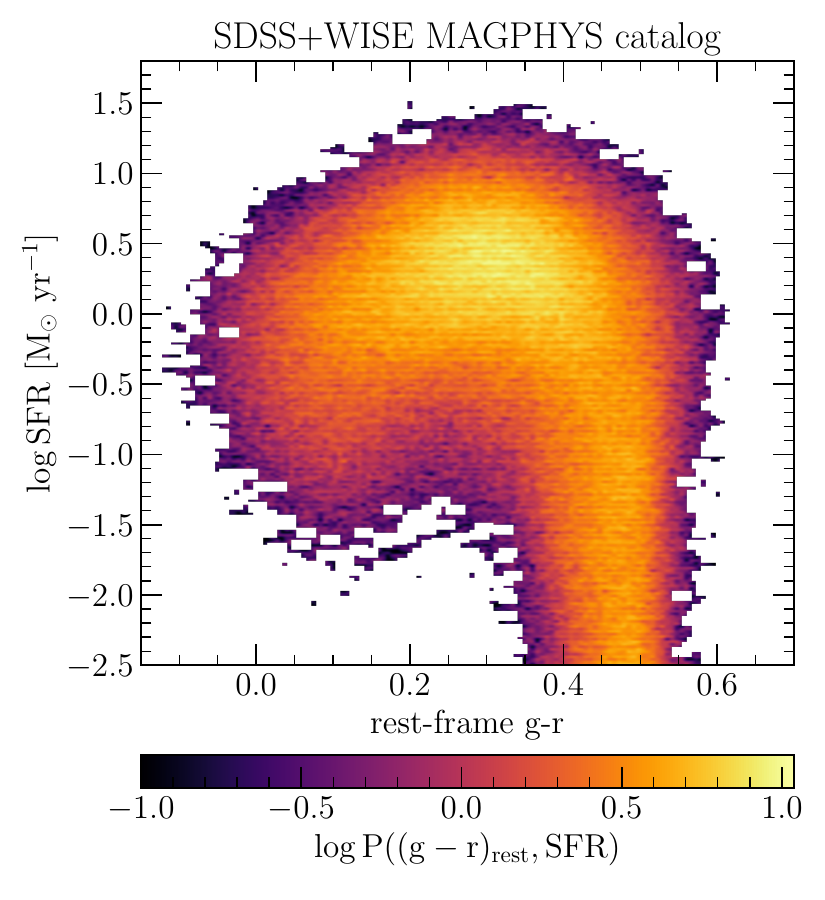}
    \caption{\textit{Left:} Redshift evolution of the stellar mass function's median $\Phi(M_\star, z)$ adopted from \cite{Leja2020}. The color shading indicates different redshift
    values. For comparison, the dashed black line denotes the median of the stellar mass function used by \cite{Bochenek2021}.
    \textit{Middle:} Hybrid-full $\mathrm{SFR-M_\star}$ probability density for $\mathrm{z=0.7}$.
    The colormap shows the conditional density field using the 
    trained normalizing flow of \cite{Leja2022} above the mass completeness threshold ($M_{comp}$; vertical dotted green line), and the
    hybrid-full prescribed (normalized) probability density $p(SFR | M_\star, z)$ below $M_{comp}$. 
    For aesthetic purposes, we have applied a mask ignoring low-probability regions. 
    The black solid line is the ridge-line best fit by \cite{Leja2022}. 
    \textit{Right:} $\mathrm{SFR}$-color probability density extracted from SDSS+WISE SED fitted data \citep{Chang2015}.
    A Gaussian filter of $\sigma=1/3$ has been applied to smooth the density field.}
    \label{fig1}
\end{figure*}

Fig.~\ref{fig1} (left panel) shows the posterior median of the stellar mass function as a function of stellar mass
for a variety of redshift values. For comparison, we also plot 
the median of the mass function (dashed black line) employed by \cite{Bochenek2021}. The slope in the low-mass range (that is, $M_\star \lesssim 10^{10}$) of the two mass functions differs significantly, thereby affecting the expected fraction of low-mass galaxies in our mock generated samples.

\subsubsection{$\mathrm{\log SFR - \log M_\star - z}$ probability density} \label{sec2.1.2}

In order to construct a population of galaxies weighted by SFR to test whether FRB host galaxies trace SF, we need a model for $p(\mathrm{SFR} | M_\star, z)$, the conditional probability distribution function (PDF) of SFR for a galaxy of given stellar mass at a given redshift. Many studies (e.g., \citealt{Elbaz2007, Kouroumpatzakis2023}) constrain the relationship between stellar mass and SFR. The $\mathrm{SFR}-M_\star$ probability density is sufficiently described 
by three components (see for example \citealt{Speagle2014}): the SFMS where galaxies are log-normally distributed
around the mode SFR value of a star-forming galaxy, a long tail of low-SFR galaxies
corresponding to the quiescent population, and a sparse population of starburst galaxies exhibiting higher SFR than SFMS galaxies \citep{Caputi2017, Rinaldi2022}.

In this work, we make use of the trained normalizing
flow presented by \cite{Leja2022}, which enables the modeling of the full probability density
function $\rho(\mathrm{SFR},M_\star,z)$ in the $\log \mathrm{SFR} - \log M_{\star} -z$ space. Thus, the PDF in SFR is given by
\begin{equation}
    p(\mathrm{SFR} | M_\star, z) = \dfrac{\rho(\mathrm{SFR},M_\star,z)}{\int \rho(\mathrm{SFR},M_\star,z) ~\mathrm{d\log SFR}}.
\end{equation}
The neural network (NN) has been
trained on parameters inferred from SED-fitting of photometric data of the COSMOS-2015 and 3D-HST UV-IR catalogs in the redshift
range $(0.2-3)$; the same dataset used to model the mass function discussed in Sect.~\ref{sec2.1.1}. The galaxy SED-fitting code used to estimate $(\mathrm{SFR}, ~ M_\star)$ values of each galaxy is
\texttt{Prospector-$\alpha$} \citep{Leja2020b}, which is built in the \texttt{Prospector} \citep{Leja2017, Johnson2021} Bayesian inference framework.

The trained NN suffers from the mass completeness threshold $M_{comp}$, due to optical selection effects and the nature of the fit. Both the COSMOS and 3D-HST surveys are magnitude-limited in the $r-$band, leading to incomplete galaxy samples below $M_{comp}$ ($\sim 10^{8.7-9.2}~M_\odot$ depending on redshift). The lack of low-mass galaxies in the training dataset of the normalizing flow means that the predictions outside of the training set are meaningless. For further details see Appendix \ref{App.SFR-Mstar}. 

To tackle this issue, we adopt the approach of \cite{Sharma2024} and construct a hybrid scheme 
consisting of the trained NN for $M>M_{comp}$ and $z>0.2$ and a prescribed probability density below $M_{comp}$, as well as at $z<0.2$,
based on the ridge-line (i.e., the SFMS) best fit of the conditional density field $\rho(\mathrm{SFR}, M_\star | z)$. 
We extrapolate the ridge-line model below the mass completeness threshold and for $z<0.2$, as well. 
The ridge-line best fit model (hereafter, ridge model) is a broken power-law fit obtained by \cite{Leja2022} 
describing the mode of $\mathrm{SFR}$ as a function of $M_\star$ for star-forming galaxies:

\begin{align}
    \log(\mathrm{SFR}_m) = \begin{cases}
    a \log(M/M_t) + c & M > M_t, \\
    b\log(M/M_t) + c & M\leq M_t,
    \end{cases}
    \label{eq.4}
\end{align}
where $a,~ b,~ c, ~\text{and} ~M_t$ are redshift-dependent quantities and are parametrized by a quadratic in redshift.
For further details, we refer the reader to \cite{Leja2022}.

While this model gives the mode of the conditional 
density field $\rho(\mathrm{SFR}, M_\star | z)$ as a function of stellar mass, 
we need the full posterior distribution in the $\mathrm{SFR}-M_\star$ plane. 
To account for the scatter around the mode, we apply the following prescription. 
The probability density $p(\mathrm{SFR} | M_\star, z)$ below $M_{comp}$ follows the full probability distribution of $p(\mathrm{SFR} |  M_\star, z)$ at $M=M_{comp}$ with an offset equal to the difference between the SFR mode values predicted by the ridge model for $M_\star$ and $M_{comp}$, respectively.
In other words, we assume that the overall shape of the PDF is preserved across mass in the low-mass regime, while shifting according to Eq.~\eqref{eq.4}.
This assumption is physically motivated as the physics governing SF and stellar feedback are expected to be pretty much self-similar in this mass range (as opposed to e.g. higher masses where galaxies start to quench) -- this means the scatter and slope do not really evolve at low mass. Below $z=0.2$, we assume that the probability density is identical to that of $z_{ref}=0.25$, but with an offset set by the difference of the ridge lines corresponding to $z_{ref}=0.25$ and z. Small variations in the spread of the prescribed PDF below $M_{comp}$ and the choice of $z_{ref}$ do not affect the results of this work. We refer to this model as the hybrid-full scheme.

In Figure \ref{fig1} (middle panel), we show the hybrid-full probability density $\rho(\mathrm{SFR}, M_\star | z)$ 
used in this work for redshift $z=0.7$. The vertical dotted green line corresponds to the mass completeness threshold. 
Above this mass value, the probability density is given by the trained NN, while below it the probability density is computed through the prescription described above. The ridge-line best fit (black solid line) by \cite{Leja2022} is also drawn. By construction the prescribed probability density follows the ridge line.

\subsubsection{Sampling function} \label{sec2.1.3}

To generate a mock galaxy sample and weight it by different physical quantities,
we write down a general sampling function, which can be used to sample
stellar masses, the distribution of which follows a desired profile, e.g.,
stellar mass function weighted by SFR. 
The sampling function is  

\begin{equation}
    P_{W}(M_\star; z) \equiv \frac{\Phi(M_\star, z) W(M_\star, z)}{\int \Phi(M_\star, z) W(M_\star, z) ~\mathrm{d} \log M_\star },
    \label{eq.sampling_func}
\end{equation}
where $W(M_\star, z)$ is a weight function
and the integral in the denominator normalizes the PDF. The integration spans stellar masses in the range $\log(M_\star / M_\odot) \in (6.5, 12.5)$; however, the choice of integration limits has no significant impact on the results. The stellar mass is sampled in logarithmic space.

In this work, we focus on three specific weight functions, but the machinery developed in this work can be employed to generate mock galaxy samples weighted by
whatever weight function is desired. The weight function options are:

\begin{align}
    W(M_\star, z) = \begin{cases}
			S(M_\star; z) & \text{if weighted by SFR}, \\
			M_\star & \text{if weighted by stellar mass}, \\
			1 & \text{unweighted},
		\end{cases}
    \label{eq_weoght}
\end{align}
where $S(M_\star, z)$ stands for an effective SFR value and has the following
functional form
\begin{equation}
    S(M_\star,z) \equiv \int \mathrm{SFR}~ p(\mathrm{SFR} | M_\star, z) ~\mathrm{d \log SFR},
\end{equation}
where $p(\mathrm{SFR} | M_\star,z)$ is shown in Fig.~\ref{fig1} (middle panel) for a given redshift (also see Sect.~\ref{sec2.1.2}). We want to weight the sample by linear $\mathrm{SFR}$, thus the integral is over $\mathrm{\log SFR}$ since $p(\mathrm{SFR} | M_\star,z)$ is by default computed in logarithmic bins. If only the ridge-line fit
is considered, then $p(\mathrm{SFR} | M_\star, z)$ is essentially a $\delta$-Dirac distribution function, i.e.,
$p(\mathrm{SFR} | M_\star, z) = \delta(\mathrm{SFR} - \mathrm{SFR}_m(M_\star,z))$. We stress that the unweighted option (i.e., $W=1$), 
in which each galaxy gets equal weight, is nonphysical and is only included in this discussion for comparison purposes between weighted and unweighted populations.

\subsubsection{Assigning an SFR value to each sample galaxy}
\label{sec2.1.4}

$P_W$ (Eq.~\ref{eq.sampling_func}) is sampled numerically to generate mock galaxy stellar mass samples.
We next assign each galaxy an SFR value by sampling the conditional density field $p(\mathrm{SFR} | M_\star, z)$ (Sect.~\ref{sec2.2.1}) if a mass-weighted or unweighted sample is desired; we sample from the distribution $\mathrm{SFR} \times p(\mathrm{SFR} | M_\star, z)$ if we want to generate an SFR-weighted population. The SFR determines the galaxy's mass-to-light ratio in our prescription. This is critical 
when comparing the sample with observational data coming from optical surveys, which
suffer from magnitude limits (see Sect.~\ref{sec2.4}).

\subsection{Generating apparent $r-$band magnitudes for mock galaxies} \label{sec2.2}

Using the prescription described above, 
we sample stellar mass and $\mathrm{SFR}$ values for each mock galaxy. 
However, we need to translate those values into directly observed quantities, 
such as the $r-$band apparent magnitude, to account for optical selection biases. 

\subsubsection{Modeling spatial distribution} \label{sec2.2.1}

In order to predict apparent magnitudes for each mock galaxy,
we first need to assign a distance to it. \cite{Sharma2024} employed a uniform distribution in redshift. 
Here, we relax this condition and test two additional spatial distributions: (1)
assigning all galaxies the mean redshift value within a specific redshift range, and (2) a uniform distribution in comoving volume. For the latter case,
we sample redshifts as follows.

Suppose the redshift range is $(z_1,z_2)$ and we wish to sample galaxy distances 
uniformly in comoving space. The PDF, in this case, is
\begin{equation}
    p(\chi) ~\mathrm{d}\chi = \dfrac{4\pi \chi^2}{V(z_1,z_2)} ~\mathrm{d} \chi,
    \label{eq.distances}
\end{equation}
where $V(z_1, z_2) = (4\pi/3) [\chi^3(z_2) - \chi^3(z_1)]$
and $\chi(z)$ is the comoving distance of a galaxy at redshift z. We sample comoving distances for each galaxy and subsequently solve for redshift using \texttt{astropy} (see also Appendix \ref{App.Distances}).

\subsubsection{Modeling the mass-to-light ratio distribution} \label{sec2.3}

To sample luminosities for each mock galaxy, we introduce another ingredient in our prescription, which is the mass-to-light ratio 
in the $r-$band, i.e., $(M_\star/L_r)_{rest}$ where the subscript $``rest"$ denotes
that this quantity is measured in the galaxy's rest frame
as opposed to $``obs"$, which refers to quantities measured in the
observed frame. 

\cite{Sharma2024} utilized estimates
from COSMOS-2015 and 3D-HST galaxy catalogs, 
accounting for variations across different redshift ranges. They considered
a log-normal distribution centered on $(M_\star/L_r)_{rest} = 1$ in solar units (where the luminosity is normalized to the Sun's bolometric luminosity) with a standard deviation
spanning from $\sigma = 0.2$ for $z\leq0.2$ to $\sigma = 0.3$ for $z\in [0.4,0.7]$. However, $M_\star / L_r$ depends strongly on the stellar population and SFR; our samples span a wide range of SFR and thus we need to account for this dependence.

As can be seen in the middle panel of Fig.~\ref{fig1}, 
there is a large fraction of galaxies that lie well below the SFMS, consisting a
population of quiescent galaxies with significantly larger $(M_\star/L_r)_{rest}$ (i.e., being fainter)
that are not represented by the assumed log-normal $(M_\star/L_r)_{rest}$ distribution. 
Therefore, a more sophisticated prescription is required to account for this effect. 
Below, we use the sampled $\mathrm{SFR}$ of each galaxy to predict a rest-frame color $(g-r)_{rest}$ (where $g_{rest},~r_{rest}$ are the rest-frame magnitudes in the $g-$band and $r-$band, respectively; Sect.~\ref{sec2.3.1}), which
then is employed to sample a $(M_\star / L_r)_{rest}$ value (Sect.~\ref{sec2.3.2}). Finally, we apply a K-correction
in the $r-$band to get the apparent magnitude (Sect.~\ref{secKr-corr}).

\subsubsection{From $\mathrm{SFR}$ to $(g-r)_{rest}$} \label{sec2.3.1}

We utilize a model which links the $\mathrm{SFR}$ of each galaxy to the $(M_\star/L_r)_{rest}$ through the rest-frame optical color $(g-r)_{rest}$. We make use of the SDSS+WISE MAGPHYS catalog\footnote{\url{https://irfu.cea.fr/Pisp/yu-yen.chang/sw.html}} released by \cite{Chang2015}, which contains SED fits to 9-band photometry of $\sim 10^6$ galaxies at $z<0.2$ to extract the conditional probability density $P((g-r)_{rest} | \mathrm{SFR})$ and sample colors for all mock galaxies. The full probability density $P((g-r)_{rest}, \mathrm{SFR})$ is shown in the right panel of Fig.~\ref{fig1}. While there exists a broad correlation between rest-frame color and $\mathrm{SFR}$, the intrinsic scatter in this relation is very large (as can be seen in Fig.~\ref{fig1}) while the quiescent galaxies lie in the high color end. Drawing a random $(g - r)$ from a probability distribution given the $\mathrm{SFR}$ for star-forming galaxies is thus a reasonable approximation, ensuring the capture of the full color distribution of star-forming galaxies while correctly distinguishing between star-forming and redder, quenched populations.

\subsubsection{From $(g-r)_{rest}$ to $(M_\star/L_r)_{rest}$} \label{sec2.3.2}

The hierarchical Bayesian model (HBM) offered by \cite{Li2022} allows us to 
sample $(M_\star/L_g)_{rest}$ for a given $(g-r)_{rest}$. This model accounts for galaxy evolution
and is redshift-dependent. It has been trained on data in the redshift range $0.5 < z < 3$. To encompass lower redshift values,
we reasonably extrapolate its validity down to redshift zero. The HBM $P((M_\star/L_g)_{rest} | (g-r)_{rest}, z)$
 is given by Eq. (1) in \cite{Li2022}. Given the color 
$(g-r)_{rest}$ and the sampled $(M_\star/L_g)_{rest}$, we
directly solve for $(M_\star/L_r)_{rest}$.

\subsubsection{$K_r$-correction} \label{secKr-corr}

FRB host galaxies lie at redshifts $z>0$, while SED-fitting codes offer
rest-frame mass-to-light ratios (see e.g., \citealt{Li2022}; Sect.~\ref{sec2.3.2}), so we need to 
apply the so-called K-correction to relate the rest-frame magnitude to the observed one \citep{Oke1968}. This
effect was not explored in \cite{Sharma2024} and we quantify it in this study.
We focus on $r-$band observations, so we need a formula for the K-correction in the $r-$band (i.e., $K_r$-correction).
We adopt the model offered by \cite{Kim2023}, which relates $K_r$ to the color $(g-r)_{rest}$,  that is, 
\begin{align}
    K_r &\equiv m_{r,rest} - m_{r,obs}  = \notag \\
     &\left[13.3(g-r)_{rest} - 0.5 - 3.5 (z-1.47)^2\right] \times \log(1+z),
     \label{eq.11}
\end{align}
where $m_{r,rest}$ ($m_{r,obs}$) denotes the rest-frame (apparent) $r-$band magnitude. This relationship was obtained by fitting simulated galaxy data in the 
redshift range $0.625 \leq z \leq 2$ and does not account for the dispersion. For the purposes of this work, we extrapolate it to lower redshift values (see also Appendix \ref{App.Kcorr}). 

The observed $(M_\star/L_{r})_{obs}$ reads 
\begin{equation}
    (M_\star/L_{r})_{obs} = (M_\star/L_r)_{rest} ~10^{-(K_r/2.5)}.
    \label{eq.12}
\end{equation}

\begin{figure*}[ht!]
    \centering
    \includegraphics[width=\textwidth]{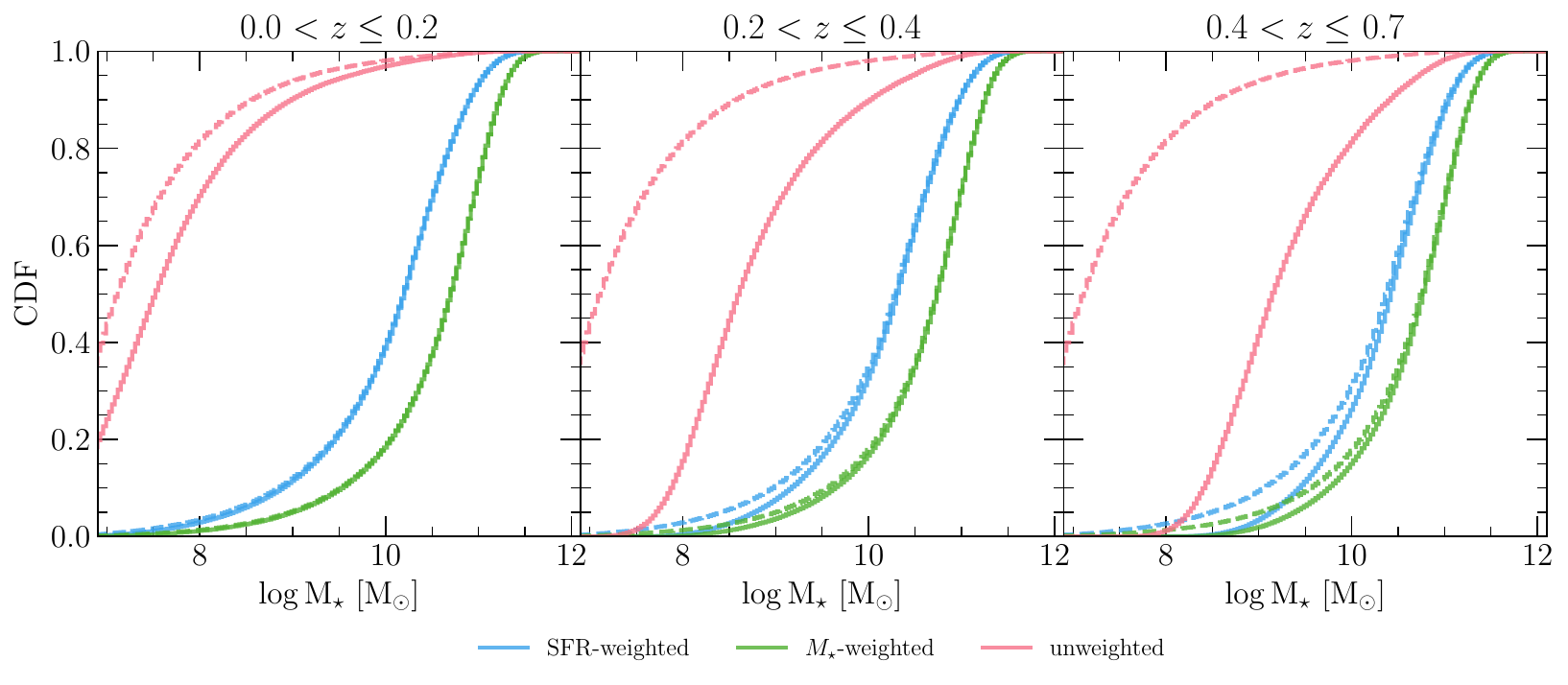}
    \caption{CDF in stellar mass of various mock galaxy samples in three different redshift ranges. The unweighted, SFR-weighted, and mass-weighted population are shown in magenta, blue, and green, respectively. Dashed lines refer to full mock galaxy samples, while the solid ones correspond to the galaxy samples after applying the magnitude limit $r=23.5$. The ridge-fit line is utilized to relate the SFR to stellar mass, while the mass-to-light ratio distribution is the one used by \cite{Sharma2024}. Distances are sampled uniformly in redshift.}
    \label{fig2}
\end{figure*}

\subsection{Handling optical selection limits} \label{sec2.4}

The sampling prescription described above yields a sample of galaxies with
realistic joint distributions of $M_\star$, $\mathrm{SFR}$, $(M_\star/L_r)_{obs}$, and redshift. However, optically fainter galaxies fall below the magnitude limit of the current optical surveys (see Sect.~\ref{sec3.1}). 
This induces selection biases towards brighter, more massive galaxies in the
sample of FRB host galaxies. To account for this 
effect in our mock generated sample, we impose an upper limit
in the apparent $r-$band magnitude $m_r^{cut}$($=23.5$ in this study) and drop all mock galaxies
that are fainter. To compute $m_{r,obs}$, we use
\begin{align}
    m_{r,obs} = {\cal M}_{r,\odot} + & 5\log(d_{L, \mathrm{Mpc}}) + 25 \notag \\
    &- 2.5 (\log M_{\star} - \log(M_\star/L_r)_{obs}),
\end{align}
where $d_{L,\mathrm{Mpc}}$ is the luminosity distance in units of $\mathrm{Mpc}$, ${\cal M}_{r,\odot}= 4.65$ \citep{Willmer2018} 
is the absolute magnitude of the sun in the $r-$band and $(M_\star/L_r)_{obs}$ includes the $K_r$-correction (see Sect.~\ref{secKr-corr}). Note that $L_r$ is normalized to $L_{r,\odot}$ in the above formula.

\subsection{Summary of prescription to generate mock host galaxy sample }
\label{sec2.5}

For each galaxy, we first assign a distance as described in Sect.~\ref{sec2.2.1}. Next, we sample its stellar mass using a weighting function to approximate the process by which different sources trace various galaxy properties (Sect.~\ref{sec2.1.3}). We then sample an $\mathrm{SFR}$ following the properly weighted probability density in the $\mathrm{SFR}-M_\star-z$ space (Sect.~\ref{sec2.1.4}). Subsequently, we assign a rest-frame optical color $(g-r)_{rest}$ based on the $\mathrm{SFR}$ (Sect.~\ref{sec2.3.1}) and utilize the probability density $P((M_\star / L_g)_{rest} | (g-r)_{rest}, z)$ (Sect.~\ref{sec2.3.2}) to sample a $(M_\star / L_g)_{rest}$ value. This is then converted to $(M_\star / L_r)_{rest}$ using the sampled $(g-r)_{rest}$ color, and we apply the $K_r$-correction to obtain $(M_\star / L_r)_{obs}$ (Sect.~\ref{secKr-corr}). Finally, we impose a magnitude cut (Sect.~\ref{sec2.4}), rejecting galaxies with an apparent magnitude of $m_r > m_r^{cut}$. This process is repeated $\cal{N}$ times to generate a sample of mock galaxies that follow the desired distributions.

\section{Cumulative stellar mass distribution of mock generated populations} \label{sec3}

In the previous section, we described how to generate mock galaxy samples
that follow specific distributions in stellar mass, SFR, color, and mass-to-light ratio, and are weighted by a physical quantity.
Here, we offer a detailed comparison of different mock galaxy populations and quantify the effect of various parts of the 
aforementioned prescription before proceeding to the comparison with data (Sect.~\ref{sec4}). The primary observable considered in this study is the cumulative distribution in stellar mass, as stellar mass measurements through SED fitting are more reliable than those of $\mathrm{SFR}$. For a comparison of the cumulative distribution in $\mathrm{SFR}$, see Appendix \ref{App.SFR_cdf}.

In what follows, we generate mock samples by constructing many realizations of a given
population to ensure sufficient sampling of the posterior distributions. We apply
a magnitude cut in the $r-$band at $m_r^{cut} = 23.5$, corresponding to the magnitude limits in the optical surveys from which the data
used in this work were obtained (see Sect.~\ref{sec4.1}). We bin the galaxy samples in three different
redshift ranges; $z\in \{(0,0.2],~(0.2,0.4], ~(0.4,0.7]\}$, following \cite{Sharma2024}. The stellar mass
range that we use is $\log (M_\star / M_\odot) \in [6.5, 12.5]$. We split the discussion into five distinct parts: (1) the differences among weighted populations (Sect.~\ref{sec3.1}), (2) the effects of the full posterior distribution in the $\mathrm{SFR}-M_\star$ plane (Sect.~\ref{sec3.2}), (3) the importance of modeling accurately the full mass-to-light ratio distribution (Sect.~\ref{sec3.3}), (4) the impact of the underlying spatial distribution of mock galaxies (Sect.~\ref{sec3.4}), and (5) the effects of the $K_r$-correction (Sect.~\ref{sec3.5}).

\subsection{Comparing weighted populations} \label{sec3.1}

To understand how the choice of weight (Eq.~\ref{eq_weoght}) affects the cumulative distribution function (CDF), we generate different samples with three weighting schemes: 
(1) unweighted, (2) SFR-weighted, and (3) mass-weighted. For simplicity and to isolate the effects of the weight factors, we employ the ridge-fit model to describe the $\mathrm{SFR}-M_\star - z$ relationship, assume a uniform distribution in redshift, and use the Gaussian model implemented in \cite{Sharma2024} for the mass-to-light ratio.
Fig.~\ref{fig2} plots the resulting CDF in stellar mass of the three weighted schemes, in the three redshift bins, both without and with the optical magnitude cut. 

At the low-redshift range only a small fraction of galaxies is filtered out by the magnitude cut, except for the unweighted sample (magenta lines), which includes many low-mass faint galaxies. At higher redshift, a larger fraction of low-mass galaxies falls below the magnitude limit. While the FRBs from these hosts can still be seen, the host galaxies themselves are not detectable.

Not surprisingly, the weighted samples (blue, green lines) favor more massive galaxies than the unweighted sample. The CDF of the mass-weighted sample rises at higher masses than does that of the SFR-weighted sample, as the sampling function is steeper when weighted by mass. All CDFs converge above $\log (M_\star /M_\odot) \gtrsim 11$, as the stellar mass distribution (Eq.~\ref{eq.2}) cuts off beyond that point.

\begin{figure*}[ht!]
    \centering
    \includegraphics[width=0.8\textwidth]{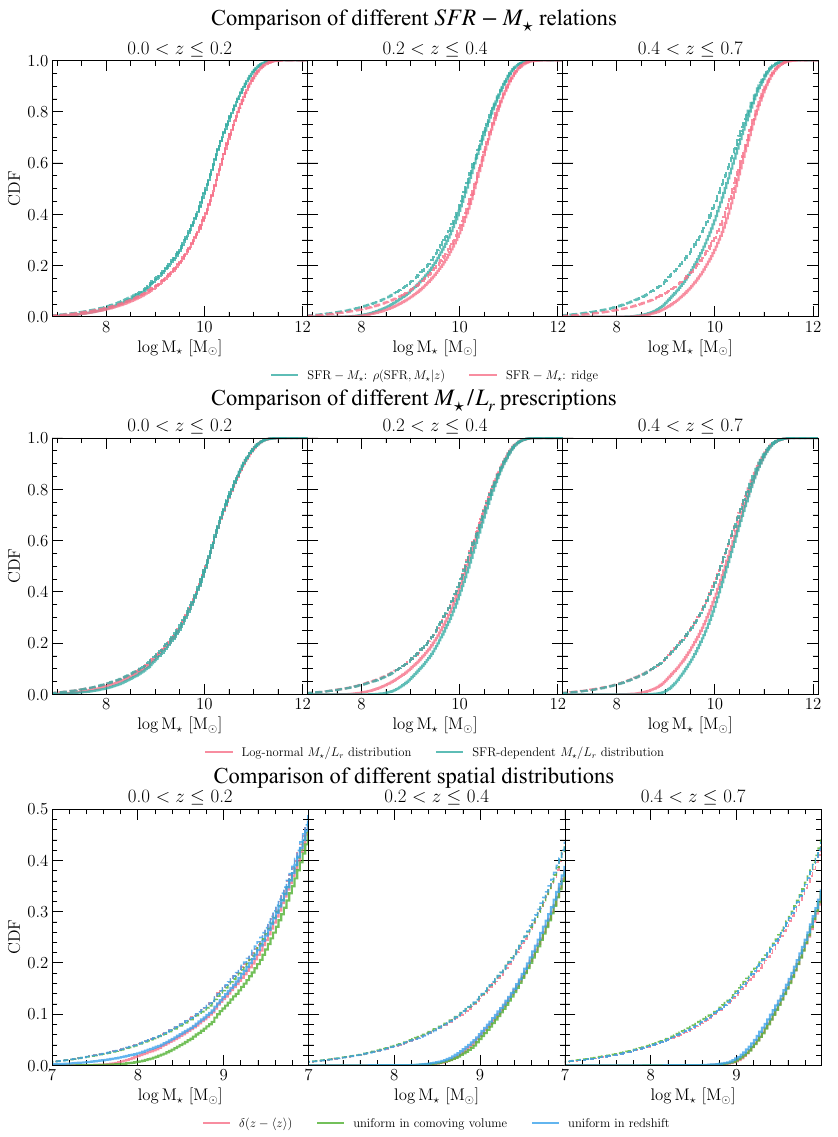}
    \caption{Solid (dashed) lines display the
(non) magnitude-limited samples. 
    \textit{Top:} Similar to Fig.~\ref{fig2}, comparing different $\mathrm{SFR}-M_\star$ relations. The population obtained using the ridge-line model is shown in magenta, while the one constructed using the full probability density $\rho(\mathrm{SFR},M_\star | z)$ is displayed in blue. The scatter in the $\mathrm{SFR}-M_\star$ plane has an impact on the distribution.
    \textit{Middle:} The galaxy population generated using a log-normal mass-to-light ratio distribution \citep{Sharma2024} is shown in magenta, while the one based on the mass-to-light ratio distribution described in (Sect.~\ref{sec2.2}) is in blue. Both populations are $\mathrm{SFR}$-weighted. The ($M_\star / L_r$) distribution can alter the fraction of low mass galaxies in the sample.
    \textit{Bottom:} Comparison of CDFs of $\mathrm{SFR}$-weighted samples using different spatial distributions. Blue, green, and magenta lines refer to samples of distances uniform in redshift, uniform in comoving volume, and assigned at the mean $z$ within each bin, respectively. 
    The spatial distribution can slightly up-shift the low mass end of the CDF when optical magnitude limits are imposed.
    }
    \label{fig3}
\end{figure*}

\subsection{Effect of $\mathrm{SFR}-M_\star-z$ relationship} \label{sec3.2}

We now demonstrate the differences in the SFR-weighted case induced by implementing the hybrid-full probability
density field $\rho(\mathrm{SFR}, M_\star | z)$ in $\mathrm{SFR}-M_\star - z$ space rather than the ridge-line fit. Fig.~\ref{fig3} (top panel) compares the CDF of the
SFR-weighted populations obtained using the ridge-line model (magenta) to that of the SFR-weighted population constructed using the full probability density (blue). 

The CDF using the hybrid-full probability distribution in $\mathrm{SFR}$ rises at lower masses than that of the ridge-line model. The implementation of the full density in $\mathrm{SFR}-M_\star-z$ increases the fraction of low-mass galaxies (with higher SFR) and down-weights the high-mass ones, leading to a shift of the CDF at lower masses.
The main reason is that the PDF in $\mathrm{SFR}$ is approximately log-normal at low masses and includes a long tail of very low (practically zero) SFR galaxies (quenched population) at high masses (see middle panel in Fig.~\ref{fig1}). This yields an increased weight of low-mass (high-$\mathrm{SFR}$) galaxies and a significantly decreased weight of the high-mass ones.

Nevertheless, the minimum mass at which the CDF is non-zero is identical between the samples with the optical magnitude cut (solid lines) across different redshift bins because the chosen mass-to-light ratio distribution is independent of the $\mathrm{SFR}$. Each galaxy is assigned a mass-to-light ratio based solely on its stellar mass and redshift, excluding any information related to $\mathrm{SFR}$, thus the minimum mass of a galaxy falling above the magnitude limit is determined by the shape of the mass-to-light ratio distribution, which does not depend on the $\mathrm{SFR}$. In this prescription, the $\mathrm{SFR}$ distribution has no effect on the sampling of apparent magnitude values when generating mock galaxies. As we demonstrate in the next subsection, though, a prescription of $\mathrm{SFR}$-dependent mass-to-light ratio can noticeably affect the predicted CDF, as well as the minimum mass at which the CDF is non-zero.

\subsection{Sensitivity to mass-to-light ratio model} \label{sec3.3}

Here we test the effect of an $\mathrm{SFR}$-dependent mass-to-light ratio, with a more general distribution than the log-normal distribution implemented by \cite{Sharma2024}. Similar to Sect.~\ref{sec3.2}, we generate two SFR-weighted populations, one based on the simplistic log-normal distribution, and the other taking into account the dependence of $M_\star/L_r$ on $\mathrm{SFR}$ (see Sect.~\ref{sec2.2}). In both samples, the full posterior distribution in the $\mathrm{SFR}-M_\star$ plane is considered and the distances are sampled uniformly in redshift. Fig.~\ref{fig3} (middle panel) shows the CDFs of both samples.

Without the magnitude cut, the CDFs of the two samples are identical and are independent of the $M_\star/L_r$ distribution. However, once the magnitude cut is implemented, the resulting populations differ significantly. The sample based on the $\mathrm{SFR}$-dependent mass-to-light ratio prescription favors more massive galaxies, shifting the CDF at higher masses. The differences are more evident at higher redshifts.
The physical explanation is the following. The $\mathrm{SFR}$-dependent mass-to-light ratio prescription allows mock quenched galaxies that lie well below the SFMS to be intrinsically fainter (i.e., be assigned a very high $M_\star / L_r$), thereby increasing the fraction of rejected low-mass galaxies due to the magnitude cut. This effect is not captured by the $\mathrm{SFR}$-independent prescription implemented in \cite{Sharma2024} in which even quenched galaxies are assigned a mass-to-light ratio from the log-normal distribution.

We note that the two CDFs overlap at higher masses, as most galaxies are bright enough to satisfy the optical selection criterion. Therefore, the mass-to-light ratio prescription mainly affects the low-mass branch of the CDF. For completeness, we present in Appendix \ref{App.samples} the distribution of all sampled physical quantities corresponding to our mock generated galaxy sample generated using the $\mathrm{SFR}$-dependent mass-to-light ratio prescription. The middle panels of Fig.~\ref{App_fig2} show the distribution of the mass-to-light ratio associated with the galaxy samples used in this section.

\subsection{Sensitivity to assumed spatial distribution} \label{sec3.4}

The spatial distribution of mock galaxies will also affect the fraction of detectable galaxies and alter the shape of the CDF. In this section, we generate three $\mathrm{SFR}$-weighted galaxy populations utilizing the full probability density in the $\mathrm{SFR}-M_\star$ plane and the $\mathrm{SFR}$-dependent mass-to-light ratio prescription, considering different spatial distributions of galaxies: uniform in redshift, uniform in volume or at a single redshift (see Sect.~\ref{sec2.2.1}). The conclusions presented in this subsection are the same for mass-weighting.

Fig.~\ref{fig3} (bottom panel) shows the resulting CDFs.  CDFs of the samples without a magnitude cut (solid lines) overlap, as neither the mass function nor the probability density $\rho(\mathrm{SFR},M_\star | z)$ evolve significantly within a narrow redshift range of $\delta z \sim 0.2$. The magnitude-selected samples (dashed lines) exhibit differences at the low-mass end that are noticeable only at low redshifts (left panel), where the luminosity distance varies rapidly with redshift. The sample that is uniform in volume has more galaxies at high redshift, which are therefore fainter and more likely to be rejected by the magnitude limit.

\subsection{Effects of $K_r$-correction} \label{sec3.5}

We have implemented the $K_r$-correction (Eq.~\ref{eq.11}) in the examples above. We find that turning this off has minimal effect on the CDFs of the weighted populations; not surprisingly, the effect is largest (but still barely noticeable) in the highest redshift bin. An explanation is provided in Appendix \ref{App.Kcorr}. The $K_r$-correction will likely play a considerable role in future surveys where FRB host galaxies may lie at higher redshifts. 

We note that the K-correction prescription used in this work is deterministic; that is, is does not include a dispersion term.  In future work, we will incorporate this effect by modeling the full posterior distribution of $K_r$ using, for instance, simulated SEDs drawn from bins of $(M_\star, \mathrm{SFR}, z)$.  However, we expect the effect to be small, as the following experiment demonstrates.  We artificially increased the $K_r$ correction amplitude of all quiescent galaxies (which are expected to be affected the most by the dispersion in the $K_r$-correction) by 10 magnitudes (i.e., we artificially decreased the observed luminosity of all quenched galaxies by four orders) and impose the magnitude limit. Even in this extraordinary scenario the resulting CDF does not differ significantly from the one where the $K_r$-correction is given by Eq.~\ref{eq.11}. This additional factor in the $K_r$ leads to a small downshift of the CDF, as quenched galaxies fall below the magnitude limit, yet the conclusions of this work are not altered. The effect is more prominent in the mass-weighted sample, as it contains a larger fraction of quenched galaxies.

\begin{figure*}[ht!]
    \centering
    \includegraphics[width=\textwidth]{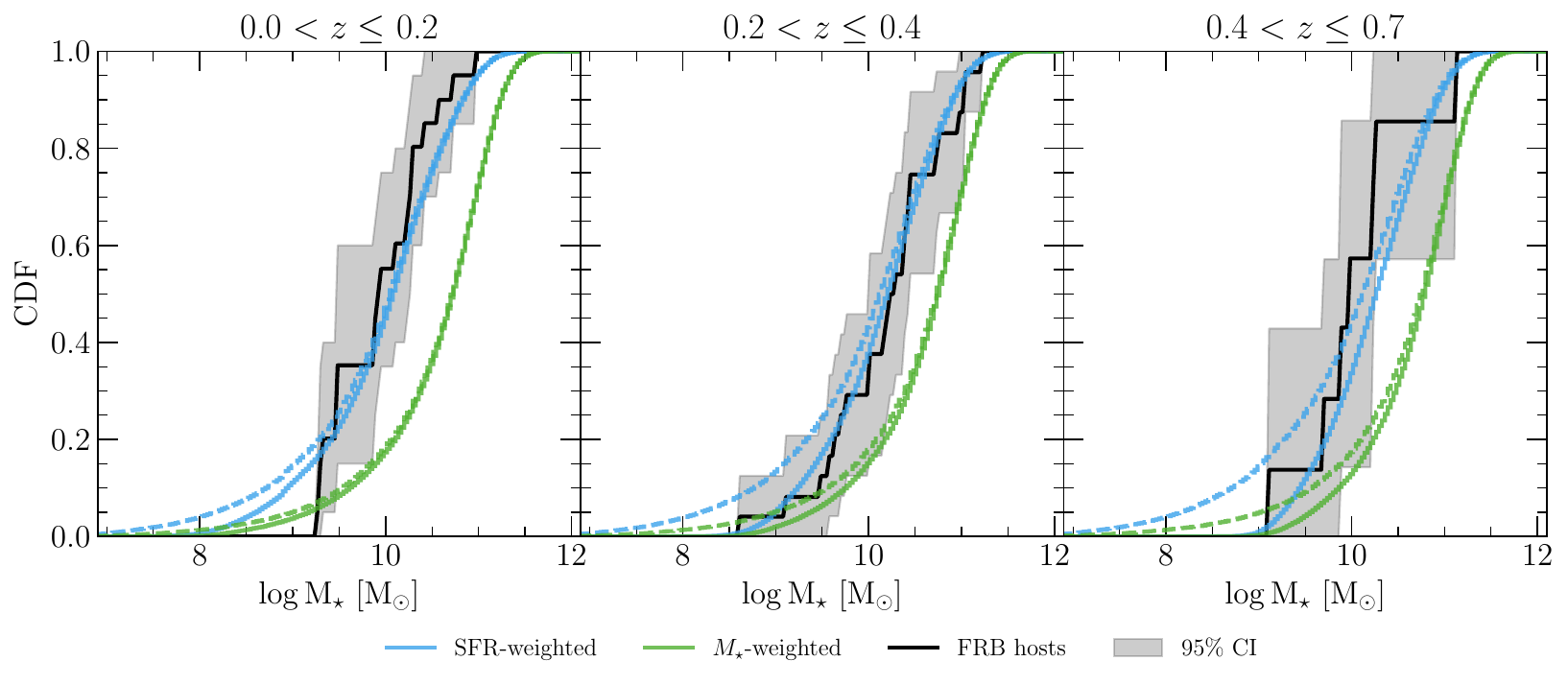}
    \caption{Comparison of mock galaxy samples tracing different quantities with FRB host galaxy data. Blue (green) line shows the CDF of the $\mathrm{SFR}$-(mass-)weighted population. Solid (dashed) lines display the (non) magnitude-limited samples. Black solid line corresponds to the FRB data (see Sect.~\ref{sec4.1}), while the gray-shaded area depicts the $95\%$ confidence interval obtained using bootstrapping. The CDFs displayed in this figure have been obtained using our prescription (Sect.~\ref{sec2.5}). Unlike the mass-weighted sample, the $\mathrm{SFR}$-weighted one is consistent with the observed data across all redshifts (see Table \ref{tab1}), indicating that FRBs trace SF. The reported deficit of low-mass hosts by \cite{Sharma2024}, in the context of FRBs as SF tracers, becomes statistically insignificant ($<1.7\sigma$).}
    \label{fig6}
\end{figure*}

\section{Comparison with FRB host galaxy data } \label{sec4}

Having presented how each component of our prescribed model (Sect.~\ref{sec2.5}) of generating mock galaxy samples shapes the CDF of the population, we proceed to compare the resulting CDF with that of publicly available data for FRB host galaxies.  

\subsection{Data sample} \label{sec4.1}

The data sample considered in this study is identical to that used in \cite{Sharma2024}. It consists of 25 FRBs, discovered by the Deep Synoptic Array (DSA-110) with secure host associations, analyzed and released by \cite{Sharma2024}, and 26 from \cite{Gordon2023} (discovered across a range of facilities) and \cite{Bhardwaj2024b} (reported in the first CHIME/FRB catalog). All FRBs considered here have an association probability ($P_{host}$) of the most likely host galaxy higher than $90\%$, as calculated with the Bayesian Probabilistic Association of Transients to their Hosts (PATH) formalism \citep{Aggarwal2021}. All host galaxies have been obtained in magnitude-limited imaging with $r-$band detection threshold of $m_r^{cut} = 23.5$. In total, there are 20 FRBs with $z \leq 0.2$, 24 FRBs with $0.2 < z \leq 0.4$, and 7 FRBs with $0.4 < z \leq 0.7$.
For an extensive discussion on the data and the procedure of identifying host galaxies we refer the reader to \cite{Sharma2024}.

For each FRB host galaxy in the sample, we have available stellar masses, $\mathrm{SFRs}$, and redshifts, measured using the \texttt{Prospector} \citep{Leja2017, Johnson2021} SED-fitting tool. In this work, we do not account for the SED-inferred errors of $M_\star$ and $\mathrm{SFR}$, which are generally small. Given the small size of the FRB host sample ($N=51$), the uncertainty in the observed CDF is primarily driven by sampling variance, which can be captured via bootstrapping. We quantify the uncertainty in the observed CDF of stellar mass by generating $10^3$ bootstrapped datasets of the same size and measure the $95\%$ confidence interval (CI). Each bootstrapped data set is obtained by sampling $N$ values from the original data set (each value can occur from $0$ to $N$ times) where $N=51$ is the data sample size.

\subsection{Comparing stellar mass cumulative distributions} \label{sec4.2}

To compare our theoretical predictions with the observed data and test whether FRBs trace either $\mathrm{SFR}$ or mass, we generate two mock galaxy samples: (1) weighted by $\mathrm{SFR}$, and (2) weighted by mass. In constructing both samples, we make use of the hybrid-full probability density field $\rho(\mathrm{SFR}, M_\star, z)$ (see Fig.~\ref{fig1}), and the $\mathrm{SFR}$-dependent mass-to-light ratio prescription (including the $K_r$-correction), and we sample galaxy distances uniformly in comoving space. 

The results are displayed in Fig.~\ref{fig6}.
The data is shown as a solid black line (representing the median), while the gray-shaded area depicts the $95\%$ confidence interval (obtained after bootstrapping the data). Blue lines refer to the SFR-weighted population, while the green ones show the mass-weighted population. The solid lines incorporate the magnitude limit, while the dashed lines do not impose a magnitude threshold.

The SFR-weighted population with the magnitude cut matches the observed CDF impressively well  across different redshift ranges. For a discussion on the sensitivity of this conclusion to the chosen stellar mass function, see Appendix \ref{App.MF}. In line with \cite{Sharma2024}, we find evidence in favor of FRB progenitors being SF tracers. In contrast, the mass-weighted population differs significantly from the data, leading to the conclusion that FRBs disfavor formation channels that trace stellar mass in host galaxies, that is, dynamic channels. 

\begin{figure*}[ht!]
    \centering
    \includegraphics[width=\textwidth]{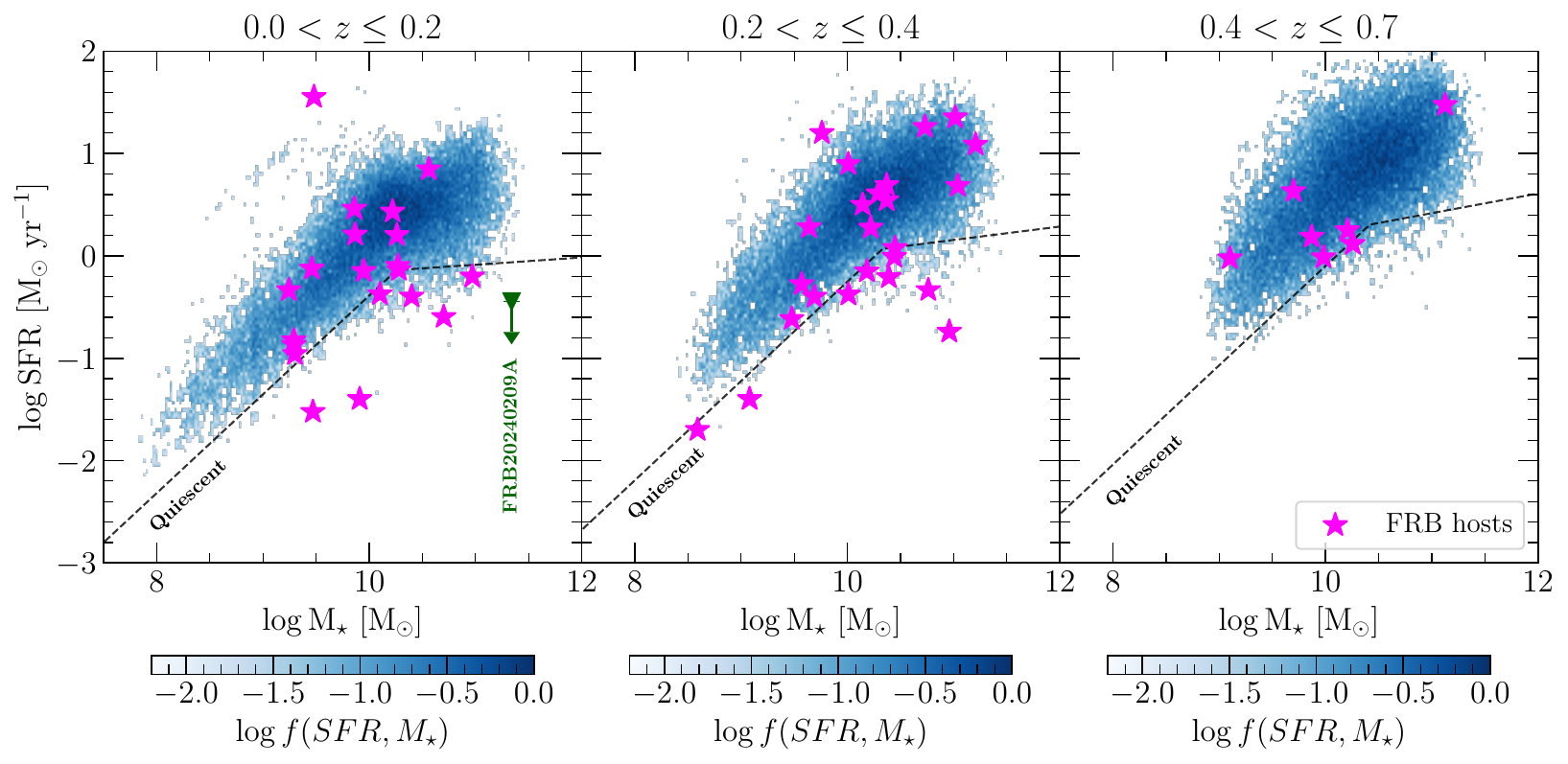}
    \caption{
    Distribution of $\mathrm{SFR}$-weighted, optically-corrected mock galaxies in the $\mathrm{SFR}-M_\star$ plane across different redshift ranges, and comparison with FRB host data. The color shading denotes the amplitude of the probability density $f(\mathrm{SFR},M_\star)$ on a logarithmic scale, while the magenta stars represent the observed FRB hosts described in Sect.~\ref{sec4.1}. The mock population shown here is $\mathrm{SFR}$-weighted, as displayed in Fig.~\ref{fig6}. The data lies in regions of the plane where the probability density of the mock population is high, indicating the consistency between the $\mathrm{SFR}$-weighted population and FRB hosts. Notice the SFR-selected sample contains a small fraction of quiescent galaxies too (falling below the dashed black line). Around the time this manuscript was ready for submission, FRB 20240209A was reported and linked to a massive, quiescent elliptical galaxy \citep{Eftekhari2024}. We include it in this figure (shown in green). }
    \label{fig6b}
\end{figure*}

\subsection{FRBs are star formation tracers} \label{sec4.3}

We make use of the so-called Kolmogorov-Smirnov (KS) statistical test to quantify the apparent agreement  between the $\mathrm{SFR}$-weighted, optically-corrected galaxy population and the observed CDF (Fig.~\ref{fig6}). We calculate the p-value in all different redshift bins, against the null hypothesis that the two distributions are drawn from the same parent distribution (Table \ref{tab1}). The computed p-values for the $\mathrm{SFR}$-weighted sample are $> 0.2$ (i.e., $\lesssim 1\sigma$) in all redshift ranges, demonstrating consistency between the FRB host galaxy sample and the mock generated, $\mathrm{SFR}$-weighted one. Conversely, the KS-test for the mass-weighted population yields significantly smaller p-values. In all redshift bins $p < 0.001$ (i.e., $\gtrsim 3\sigma$), except in the last one (i.e., $0.4 < z \leq 0.7$) where $p = 0.0018$ -- this is most likely because of the small number of FRB hosts at this range. 

These findings suggest that FRBs are consistent with tracing SF, but are not consistent with being stellar mass tracers, although a mixed population remains a plausible scenario (see Sect \ref{sec4.4}). We note that our p-values are higher for the $\mathrm{SFR}$-weighted population than those obtained by \cite{Sharma2024}. This is a consequence of our more detailed model: the implementation of an $\mathrm{SFR}$-dependent mass-to-light ratio, and sampling distances uniformly in space, decreases the fraction of the expected optically detectable low-mass galaxies, shifting the low-mass end of the CDF closer to the data. 

\cite{Sharma2024} reported a significant $(>3\sigma)$ deficit of low-mass FRB hosts at $z<0.2$ relative to SFR-selected, optically-corrected mock galaxies. Our modeling suggests that in the FRBs-as-SFR-tracers scenario, only \(\sim15\%\) of mock FRB hosts are expected to reside below $\sim10^{9.2} ~\rm{M_\odot}$ at $z<0.2$ (see Fig. \ref{fig6}). In a sample of $N=20$ (the sample size in the $z<0.2$ redshift bin), this implies an expectation of \(\sim3\) low-mass FRB hosts; however, none are observed. Given Poisson statistics, this corresponds to a $\sim1.65\sigma$ deviation—much smaller than the \cite{Sharma2024} result and therefore statistically less significant.

\begin{deluxetable}{cccc}[t!]
\tablenum{1}
\tablecaption{p-values based on KS-statistic for FRB host galaxies sample.\label{tab1}}
\tablewidth{0pt}
\tablehead{
\colhead{Model} &
\colhead{} & \colhead{Redshift} & \colhead{} \\
\colhead{Weight} & \colhead{$0 < z \leq 0.2$} & \colhead{$0.2 < z \leq 0.4$} & \colhead{$0.4 < z \leq 0.7$}}
\startdata
$\mathrm{SFR}$ & 0.47 & 0.93 & 0.23 \\
Mass & $3.4 \times 10^{-6}$ & $5.6 \times 10^{-5}$ & $1.8 \times 10^{-3}$ \\ 
\enddata
\end{deluxetable}

Fig.~\ref{fig6b} shows the distribution of mock-generated, $\mathrm{SFR}$-weighted, magnitude-limited galaxies in the $M_\star - \mathrm{SFR}$ plane across different redshift bins, overlaid with the FRB host galaxy data (magenta stars). The color shading denotes the amplitude of the probability density $f(\mathrm{SFR}, M_\star)$, smoothed for visualization purposes using a Gaussian kernel. The observed FRB hosts mostly lie in the regions where the probability density is high, demonstrating further the consensus between the mock galaxy populations and the observed one. This observation consists another piece of evidence in favor of the scenario in which FRBs trace SF. The projection of these distributions on the mass axis is shown in Fig.~\ref{fig6} (as CDFs), while the equivalent on the $\mathrm{SFR}$ axis is given in Fig.~\ref{App_fig1} (see Appendix \ref{App.SFR_cdf}). 

The vast majority of the mock galaxies lie in the SFMS given the weighting by $\mathrm{SFR}$, but a small fraction, $\lesssim 6\%$, resides below it, assuming a SFMS width of $\sim 0.5~ \mathrm{dex}$. Likewise, the majority of FRB hosts is distributed around the SFMS. Some FRB hosts lie well outside the SFMS, in particular in the low SFR, high $M_\star$ tail of the distribution. We have seen that we can explain this population of low SFR hosts even though we are weighting by SFR; a sample of $\mathrm{SFR}$-weighted potential FRB host galaxies would also contain quiescent, massive galaxies. Thus, the discovery of a small fraction of FRBs in quiescent galaxies does not rule out the single, short time-scale formation channel, nor does it serve as definitive evidence for dynamic formation pathways, as suggested by \cite{Sharma2023}. However, the detection of an FRB in a globular cluster (e.g., \citealt{Kremer2021}) supports the dynamic formation scenario, as both star formation and the total stellar mass in such environments are negligible.

The green triangle in Fig.~\ref{fig6b} is FRB 20240209A, which is associated with a massive, quiescent elliptical galaxy \citep{Eftekhari2024}. It lies at slightly higher stellar mass and lower SFR than the predicted probability density, which may result from systematic biases arising from differences in the IMF and stellar population synthesis models used in the SED fitting of this FRB’s host and those applied in modeling the stellar mass function and $\rho(\mathrm{SFR}, M_\star, z)$ utilized in this study.

Here, we only provide an upper limit ($\sim 6 \%)$ on the expected number of FRB hosts below the SFMS. The SFR is often poorly measured in quenched galaxies, as typical SFR indicators in such systems are swamped by non-SFR processes \citep[e.g.,][]{Fumagalli2014}. Thus, we interpret the $\sim 6 \%$ of galaxies below the SFMS as an upper limit.
Their SFRs could be considerably lower, suppressing the fraction of quiescent galaxies in the SFR-weighted mock sample. Precise SFR measurements in quiescent galaxies is critical to provide robust constraints on the expected fraction of SFR-selected galaxies lying below the SFMS and increase the constraining power of this approach to discerning diverse formation channels. If future FRB host samples show a  fraction of quenched galaxies significantly larger than $\sim 6 \%$, this would be strong evidence for the existence of dynamic formation channels.

\begin{figure*}[ht!]
    \centering
    \includegraphics[width=\textwidth]{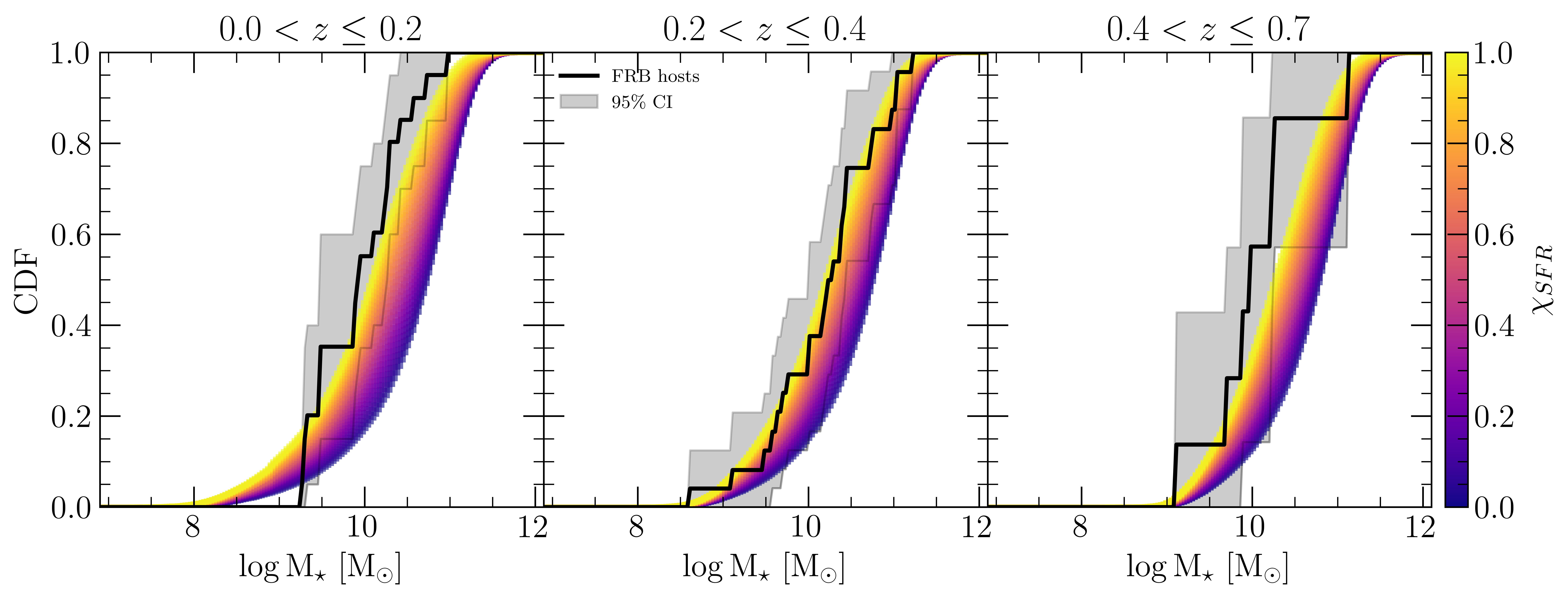}
    \caption{Similar to Fig.~\ref{fig6}. Comparison of mixed populations with the FRB host galaxy data. The left-most (right-most) colored line in each panel corresponds to the $\mathrm{SFR}$-weighted (mass-weighted) population, while those in between represent mixed populations, as indicated by the color scale; $\chi_{\mathrm{SFR}}$ is the fraction of $\mathrm{SFR}$-weighted galaxies. The black solid line and the gray shaded area, representing the observed sample, are identical to those in Fig.~\ref{fig6}. CDFs displayed in this figure have been obtained using our prescription (Sect.~\ref{sec2.5}). All lines show mock populations with the magnitude limit imposed.}
    \label{fig7}
\end{figure*}

\subsection{What percentage of FRBs tracing stellar mass is still viable?}
\label{sec4.4}

Thus far, we dealt with single formation channel models that trace either SF or stellar mass. Nevertheless, multiple formation channels have been proposed to explain the diversity of FRB properties. If so, the FRB population may include a mixture of young and old stellar progenitors.

In this exercise, we quantify the maximum allowed percentage of FRBs tracing stellar mass in a multiple formation channel model. 
To construct a mixed mock galaxy population, we combine the $\mathrm{SFR}$-weighted and the mass-weighted populations discussed in Sect.~\ref{sec4.3}. The fraction of galaxies tracing $\mathrm{SFR}$ in the mixture is $\chi_{\mathrm{SFR}}$. The results are given in Fig.~\ref{fig7}; all results incorporate the magnitude limit. 

\begin{figure}[ht!]
    \centering
    \includegraphics[width=\linewidth]{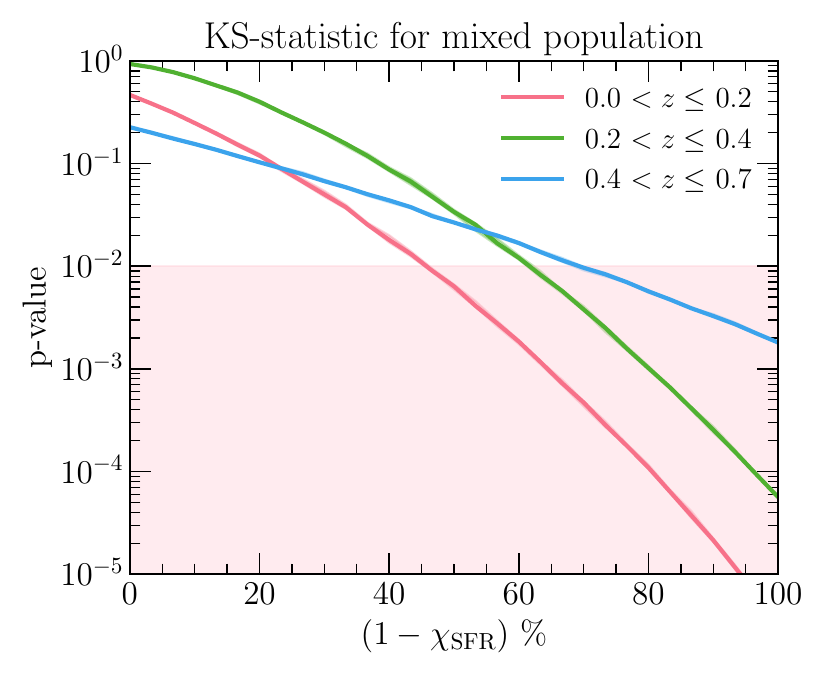}
    \caption{KS-statistic corresponding to the comparison of FRB host galaxy data with various mixed populations. ($1-\chi_{\mathrm{SFR}}$) is the fraction of the population traced by stellar mass. Each colored line represents the p-value for a different redshift bin. The pink-shaded region depicts the region with $p < 0.01$ (i.e., $\gtrsim 2 \sigma$). }
    \label{fig8}
\end{figure}

As can be seen in this figure, even a $50-50$ mixture of the two populations can still be consistent with the observed CDF. We quantify this in Fig.~\ref{fig8}, which shows the p-value as a function of the mass-weighted fraction of galaxies (i.e., $1-\chi_{\mathrm{SFR}}$) for each redshift bin. In all cases, the p-value drops with the fraction of mass-weighted galaxies in the mixture, indicating that the most favored solution has $\chi_{\mathrm{SFR}} = 1$. Nevertheless, the data is still (within $\sim 2 \sigma$) consistent with a mixed population consisting of $\sim 45\%$, $\sim 60\%$, $\sim 67\%$ mass-weighted galaxies in the redshift range $0 < z \leq 0.2$, $0.2 < z \leq 0.4$ , $0.4 < z \leq 0.7$, respectively..  
The slope of the (blue) curve corresponding to the highest redshift is smaller given the small sample size, allowing for larger deviations between the model CDF and the observed one. Increasing the number of FRBs with known host galaxies will significantly improve the constraining power of this method.

Considering a mixture of young and old progenitors, we find that while most FRBs trace SF, a significant fraction--up to $\sim 20\%$ ($50\%$) within $\sim 1\sigma$ ($2\sigma$)--can still trace stellar mass instead, while remaining consistent with the current data. This is in line with the results of \cite{Mo2025} where a mixture of populations with $\sim 60\%$ of FRBs tied to SF can explain the joint distribution of dispersion measure and scattering time in the CHIME/FRB catalog.

\section{Conclusions} \label{sec5}

In this work, we built a toolkit suitable 
for generating mock galaxy samples tracing various physical quantities that can be used to compare with future FRB host galaxy data in a relatively straightforward way. 
We made use of the FRB host galaxy dataset considered in \cite{Sharma2024} and focused on the stellar mass cumulative distribution, taking into account the magnitude limit of the host galaxy identification.
We provided a prescription
of sampling $\mathrm{SFR}$-dependent mass-to-light ratios for each mock galaxy to estimate its
apparent $r-$band magnitude, while incorporating the $K_r$-correction. In addition, we considered a more realistic spatial distribution of galaxies than did \cite{Sharma2024}; a uniform distribution in comoving volume. We compare our models with the data performing a KS test on the CDF of stellar mass. Furthermore,
we constrain the fraction
of FRBs tracing SF relative to those tracking stellar mass.  

The main findings of this study are as follows:

\begin{itemize}

    \item[1.] In line with \cite{Sharma2024}, our analysis suggests that FRBs are SF tracers, provided there is only one formation channel (Sect.~\ref{sec4.2}; \ref{sec4.3}). We generated two mock galaxy populations: (1) $\mathrm{SFR}$-weighted, and (2) mass-weighted, and found that the $\mathrm{SFR}$ one matches the CDF of FRB hosts impressively well. The KS-statistic yielded p-values significantly larger than $0.2$, verifying the consistency between the two populations across different redshift ranges. In contrast, the mass-weighted population differs remarkably from the data, making it a highly improbable scenario. We emphasize that our p-values are higher than the ones obtained from \cite{Sharma2024} as regards the $\mathrm{SFR}$-weighted population, since both the $\mathrm{SFR}$-dependent mass-to-light ratio prescription and the uniform in comoving volume spatial distribution of galaxies shift the CDF to higher masses, bringing the low mass end closer to the data, and thus alleviating the tension that was raised by \cite{Sharma2024}.~While \cite{Sharma2024} reported a significant deficit in the number of low-mass FRB host galaxies, we found this difference was only $\sim 1.65\sigma$ given our modeling of the SFR-selected galaxy population.

    \item[2.] While it has been suggested that the existence of quiescent, massive FRB hosts indicates that there must be an additional formation channel tied to stellar mass \citep{Sharma2023, Eftekhari2024}, our analysis demonstrates that $\mathrm{SFR}$-selected galaxies can still encompass a small fraction of galaxies below the SFMS (see Fig.~\ref{fig6b}).

    \item[3.] However, we have not ruled out a formation channel tied to stellar mass. We allow for a mixed population--where a fraction of FRBs trace $\mathrm{SFR}$ and the rest track stellar mass--we found that, while the $\mathrm{SFR}$-weighted population maximizes the p-value in a KS-test, a mixed population comprising up to $\sim 45\%$, $\sim 60\%$, and $\sim 67\%$ mass tracers at redshift ranges $0 < z \leq 0.2$, $0.2 < z \leq 0.4$, and $0.4 < z \leq 0.7$, respectively, can still be within $2\sigma$ (p-value $> 0.01$) of the data (see Sect.~\ref{sec4.4}). These findings suggest that FRBs may originate from both old and young progenitors, making it difficult to rule out either population with high confidence.

    \item[4.] Including the scatter in the $\mathrm{SFR}-M_\star$ plane (rather than considering only the SFMS) is particularly important when sampling $\mathrm{SFR}$-selected mock galaxies (Sect.~\ref{sec3.2}). The full posterior distribution in $\mathrm{SFR}$ gives greater weight to low-mass galaxies with high $\mathrm{SFR}$, as the high-mass end of the distribution contains a substantial fraction of quenched galaxies (well below the SFMS), which are suppressed due to their significantly low $\mathrm{SFR}$ values.

    \item[5.] The shape of the mass-to-light ratio distribution plays an important role in the comparison mock galaxy populations with data extracted from magnitude-limited host-galaxy samples. Applying a more careful modeling of the overall distribution, allowing for high $(M_\star / L_r)$ tails (associated with old, massive galaxies), yields a lower fraction of low-mass galaxies in the optically-selected sample, shifting the CDF to higher masses (see Sect.~\ref{sec3.3}). This effect becomes more important at higher redshifts where a larger fraction of low-mass (faint) galaxies is not detectable. 
    
    \item[6.] The assumed spatial distribution of galaxies does not significantly alter the mock generated galaxy sample before applying the magnitude limit, but it can modulate the predicted CDF with the magnitude limit, as the apparent magnitude is sensitive to the assumed distance of each galaxy (see Sect.~\ref{sec3.4}).

\end{itemize}

The sample of FRB host galaxies is expected to grow rapidly in the near future. Upcoming radio telescope arrays, such as the 
CHIME/FRB outriggers \citep{Lanman2024}, CHORD \citep{CHORD}, DSA-2000 \citep{Hallinan2019},
SKA Observatory \citep{Braun2015,SKA1}, and the FAST Core Array \citep{FAST-CA}, will localize thousands of FRBs in real time with unprecedented precision. Expanding the sample and accessing fainter magnitudes would significantly enhance the constraining power of this method and increase the sensitivity of the CDF of stellar mass at lower masses. Fainter magnitudes would also enable a deeper probe of the stellar mass function and the $\mathrm{SFR}-M_\star$ scatter well below the mass completeness threshold $M_{comp}$; currently set at $\sim 10^9 ~M_\odot$, thereby reducing uncertainties associated with the prescribed $\mathrm{SFR}-M_\star$ probability density. However, this is currently impractical due to the imposed magnitude limit. Moreover, obtaining precise measurements of SFR in quiescent galaxies would improve the constraints related to single formation channels (see Sect.~\ref{sec4.3}). As demonstrated in Sect.~\ref{sec4.4}, stronger constraints on the fraction of FRBs tracing stellar mass--within a multiple formation channels framework--would also become feasible with a larger dataset.

Furthermore, identifying FRB hosts at higher redshifts, where the SF density is greater, would provide additional support for the hypothesis that FRB progenitors predominantly trace SF, reinforcing the short time-scale channel and thus the young magnetar formation scenario via the CCSN pathway. However, assembling a complete FRB host sample at higher redshifts remains challenging, as it requires improved $r-$band sensitivity (i.e., a higher $m_r^{cut}$). 

The machinery offered in this study serves as a straightforward approach of comparing future FRB data with mock generated galaxy samples. The careful modeling employed in our code enables robust comparisons with FRB hosts even at higher redshifts where effects, such as the K-correction, play an important role. We encourage the research community to take advantage of our publicly available \texttt{GALFRB} (GALaxy properties for FRB progenitor identification) code to obtain more stringent constraints towards the origin of FRBs, once a larger number of FRBs has associated host galaxies.

\begin{acknowledgments}
NL would like to thank Mohit Bhardwaj, Bingjie Wang, and Kritti Sharma for fruitful discussions on various aspects of this work. NL is indebted to Ben Margalit and Asaf Horowicz for their careful reading of the manuscript and their valuable input on the proper sampling of SFRs.
NL is grateful to Christian Kragh Jespersen for insightful discussions on the machinery of the SED-fitting methods. NL also acknowledges financial support from Princeton University. NL wishes to thank the ``Summer School for Astrostatistics in Crete" for providing training on the statistical methods adopted in this work.
\end{acknowledgments}

\software{Astropy \citep{2013A&A...558A..33A,2018AJ....156..123A},
matplotlib \citep{Caswell2018},
scipy \citep{2020SciPy-NMeth},
numpy \citep{harris2020array}
}

\appendix

\section{Probability density field of the trained normalizing flow below the mass completeness threshold}
\label{App.SFR-Mstar}

In order to sample $\mathrm{SFR}$ values for the mock-generated galaxy population, we used the trained normalizing flow model for the probability density $\rho(\mathrm{SFR},M_\star,z)$ offered by \cite{Leja2022}. This model becomes highly uncertain below the mass completeness threshold $M_{comp}$. In Sect.~\ref{sec2.1.2}, we prescribed the probability density at low masses in a way that it tracks the extrapolated ridge-line best fit (i.e., the SFMS) rather than naively implementing the normalizing flow across the whole mass range. In this appendix, we justify why prescribing the probability density at low masses is necessary. 

\begin{figure}[h!]
    \centering
    \includegraphics[width=0.5\linewidth]{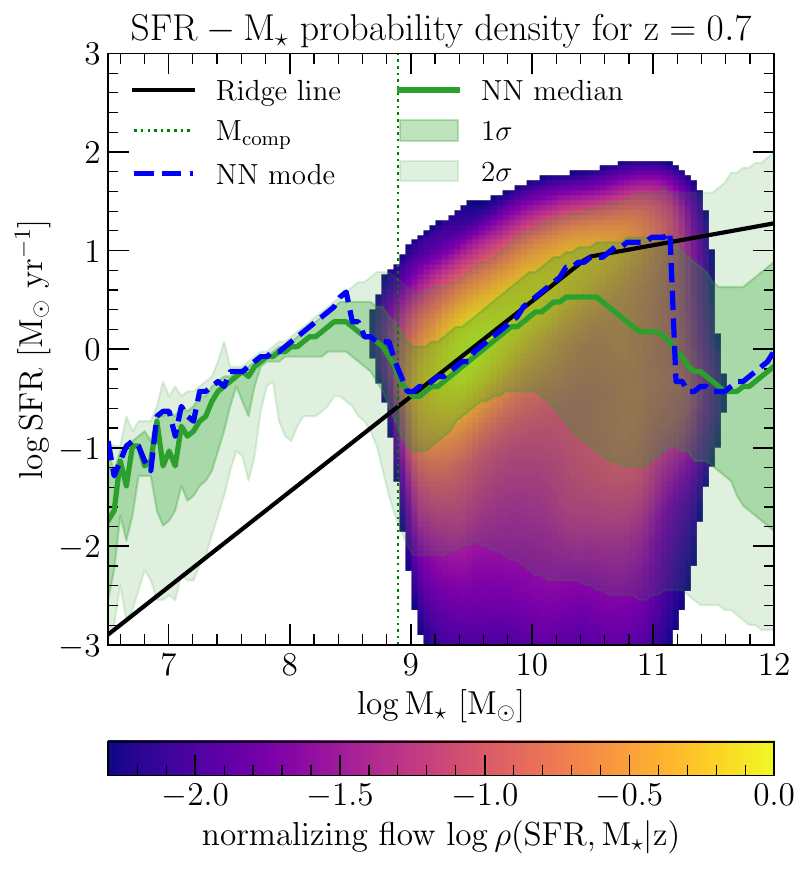}
    \caption{$\mathrm{SFR}-M_\star$ probability density field for $\mathrm{z=0.7}$.
    The color shading corresponds to the conditional density field obtained by the 
    trained normalizing flow of \cite{Leja2022}. For visualization purposes, we do not draw very low-probability regions.  
    The black solid line is the ridge-line best fit by \cite{Leja2022}. 
    The green solid line depicts the $\mathrm{SFR}$ median as a function of stellar mass computed
    using the trained normalizing flow, while the light-green strips refer to the $\mathrm{1\sigma}$ and $\mathrm{2\sigma}$
    of the PDF. The blue dashed line shows the mode (ridge) of the NN conditional probability density. The green vertical dotted line denotes the mass completeness threshold.}
    \label{App_fig2}
\end{figure}

Fig.~\ref{App_fig2} shows the normalizing flow conditional probability density $\rho(\mathrm{SFR},M_\star | z)$ for $z=0.7$. The median and mode values of the normalizing flow's probability density behave well above the mass completeness threshold and below the high mass limit where the probability density is low. However, below $M_{comp}$ both the ridge line and the median deviate from what is expected from the extrapolated best ridge-line fit, leading to an overestimation of $\mathrm{SFR}$, which would bias our samples towards high star-forming, low-mass galaxies. The trained NN under-performs below $M_{comp}$, predicting much higher SFR values than expected. It thus cannot be used in our analysis.

\section{Sampling redshifts uniformly in comoving space} \label{App.Distances}

The comoving distance of a galaxy at redshift z is given by
\begin{equation}
    \chi (z) = \dfrac{c}{H_0} \int_0^z \dfrac{\mathrm{d}z}{H(z)/H_0}, 
    \label{comoving}
\end{equation}
where $c$ is the speed of light in vacuum, $H(z)$ is the Hubble function, and $H_0$ is Hubble's constant.
To sample a distance, we draw a random number $\xi_i$ between 0 and 1 and solve the following equation 
\begin{equation}
    \xi_i = \int_{\chi(z_1)}^{\chi(z_i)} p(\chi) ~ \mathrm{d}\chi,
\end{equation}
which yields
\begin{equation}
    \xi_i = \dfrac{[\chi(z_i)/\chi(z_2)]^3 - [\chi(z_1)/\chi(z_2)]^3}{1 - [\chi(z_1)/\chi(z_2)]^3},
\end{equation}
where $\chi(z_i)$ is the comoving distance corresponding to the random number $\xi_i$ and $p(\chi)$ is given by Eq.~\eqref{eq.distances}. 
Inverting the above expression, we get
\begin{equation}
    \chi(z_i) = \chi(z_2) \sqrt[3]{[\chi(z_1)/\chi(z_2)]^3 + \xi_i\left(1 - [\chi(z_1)/\chi(z_2)]^3\right) },
\end{equation}
and subsequently we solve for redshift using Eq.~\eqref{comoving}, i.e., $\chi(z_i) \to z_i$ (\texttt{astropy}).

\section{Mock generated galaxies: distributions of various sampled physical quantities} \label{App.samples}

\begin{figure}[ht!]
    \centering
    \includegraphics[width=0.8\linewidth]{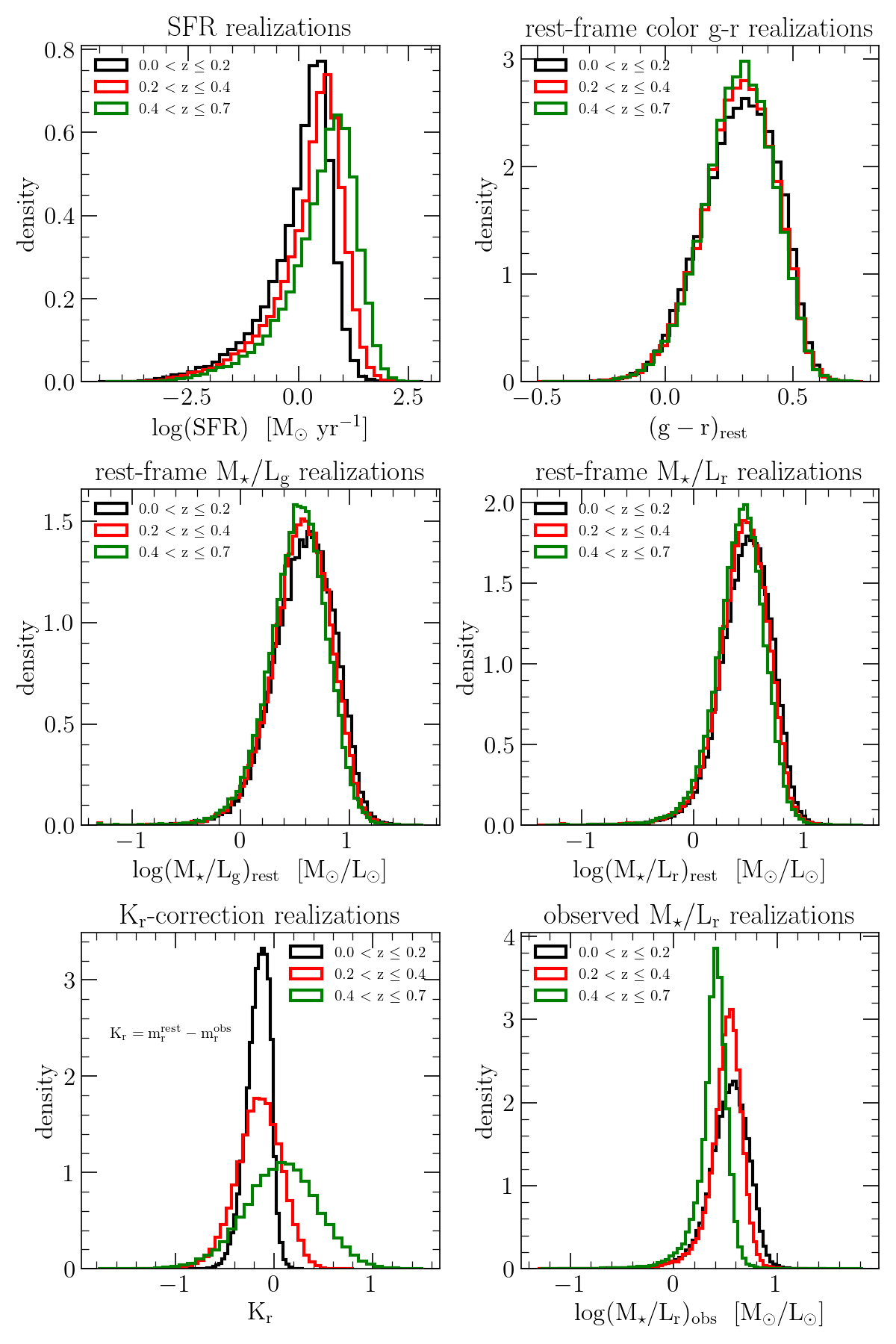}
    \caption{Various distributions of sampled physical quantities corresponding to mock generated, $\mathrm{SFR}$-weighted galaxies presented in Sect.~\ref{sec3.3}. Different colors refer to different redshift ranges.}
    \label{App_fig3}
\end{figure}

The prescription described in the main text relates $\mathrm{SFR}$ to $(M_\star/L_r)_{obs}$ by taking advantage of the $\mathrm{SFR}-(g-r)_{rest}$ relationship, as well as the $(g-r)_{rest} - (M_\star/L_g)_{rest}$ correlation and the $(g-r)_{rest} - K_r - z$ relation. As a result, we obtain joint distributions of $M_\star$, $\mathrm{SFR}$, $(g-r)_{rest}$, $(M_\star/L_g)_{rest}$, $(M_\star/L_r)_{rest}$, $K_r$, and $(M_\star/L_r)_{obs}$.
Fig.~\ref{App_fig3} shows the distributions of these quantities when using the $\mathrm{SFR}$-weighted population discussed in Sect.~\ref{sec3.3}, and the $\mathrm{SFR}$-dependent mass-to-light ratio prescription. The distribution of $\mathrm{SFR}$ (top left panel) shifts to higher values with increasing redshift, which reflects the fact that star formation increases with redshift due to cosmic evolution. It peaks at high $\mathrm{SFR}$ values, owing to the fact that galaxies have been weighted by $\mathrm{SFR}$. The distribution of color $(g-r)_{rest}$ is displayed in the top right panel.

The middle left and right panels show the distribution of $(M_\star/L_g)_{rest}$ and $(M_\star / L_r)_{rest}$, respectively. The part of the distribution where the mass-to-light ratio is around 1 corresponds to the blue/star-forming galaxies, while the high mass-to-light ratio part includes the red/quiescent galaxies. The low redshift bin includes a slightly larger fraction of quiescent galaxies, leading to a small shift of the peak at higher mass-to-light ratios. The bottom left and right panels display the distribution in $K_r$ and $(M_\star / L_r)_{obs}$, respectively. The spread in $K_r$ is small at the low redshift bin, but increases with redshift, reaching values of $\sim -1,1$ at the high redshift regime. The shape of the observed mass-to-light ratio depends on redshift, with the highest redshift population of galaxies having lower mass-to-light ratios. This contrasts with the assumption made by \cite{Sharma2024} that galaxies are log-normally distributed around $(M_\star/L_r)_{obs} = 1$. Our prescription yields fainter galaxies on average (see also Sect.~\ref{sec3.3}). 

\section{K-correction samples} \label{App.Kcorr}

\begin{figure}[ht!]
    \centering
    \includegraphics[width=\linewidth]{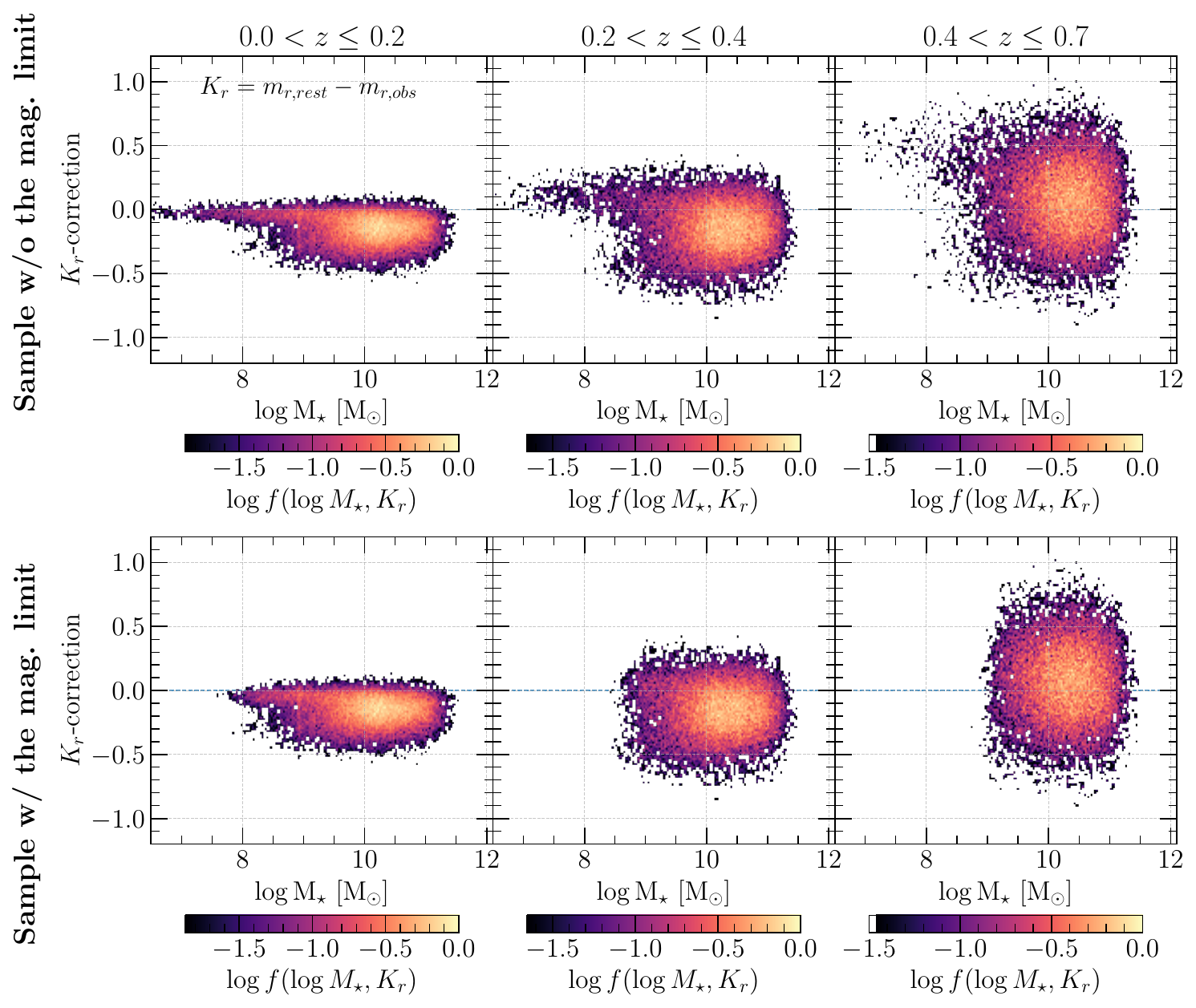}
    \caption{$K_r - \log M_\star$ probability density of the SFR-weighted mock galaxy sample used in Sect.~\ref{sec4.2}. Top (bottom) row panels show the sample without (with) the imposed magnitude limit. Each column corresponds to a specific redshift bin.}
    \label{App_fig5}
\end{figure}

In Sect.~\ref{sec3.5}, we argued that the K-correction does not play an important role in this study. Here, we employ the sample of the SFR-weighted population discussed in Sect.~\ref{sec4.2} to offer an explanation of why the K-correction does not alter the shape of the predicted CDF. Fig.~\ref{App_fig5} displays the distribution of the galaxy sample in the $K_r - \log M_\star$ plane for different redshift bins, with (bottom panels) and without (top panels) the magnitude limit. 
The full sample contains star-forming galaxies, extending down to low masses (narrow strip), and the quiescent population lying at higher masses. The spread in the $K_r$-correction increases with redshift, and the distribution projected on the $K_r$ axis is asymmetric and differs across stellar mass and redshift. 

$|K_{r}|$ is smaller than $\sim 0.4$, $\sim 0.6$, and $\sim 0.8$ in the redshift range $0< z \leq 0.2$, $0.2< z \leq 0.4$, and $0.4< z \leq 0.7$, respectively. Thus, the ratio of $(M_\star / L_r)_{rest}$ to $(M_\star / L_r)_{obs}$ due to the K-correction is
\begin{align}
    \dfrac{(M_\star / L_r)_{rest}}{(M_\star / L_r)_{obs}} \leq  10^{(|K_r| / 2.5)}   \lesssim \begin{cases} 1.4 & \text{if } 0 < z \leq 0.2, \\
    1.7 & \text{if } 0.2 < z \leq 0.4, \\
    2.1 & \text{if } 0.4 < z \leq 0.7, 
    \end{cases} 
    \label{fraction_diff}
\end{align}
where we made use of Eq.~\eqref{eq.12}. This implies that, in the context of the imposed magnitude limit, the K-correction can affect galaxies within a fractional mass range $\Delta M_\star / M_\star \lesssim 2$ only, and therefore its effect on pushing galaxies fainter than $m_r^{cut} = 23.5$ is very limited over the whole sample. The combination of this effect and the fact that the samples weighted either by SFR or stellar mass mostly contain high-mass galaxies, naturally explains the conclusion that the K-correction does not affect the predicted CDFs in this redshift range. If however, we include the dispersion in $K_r$ allowing for significantly higher values, then indeed the K-correction will modulate the predicted CDF. Such an exercise is beyond the scope of this work, and will be explored in a future paper.

\begin{figure*}[h!]
    \centering
    \includegraphics[width=\textwidth]{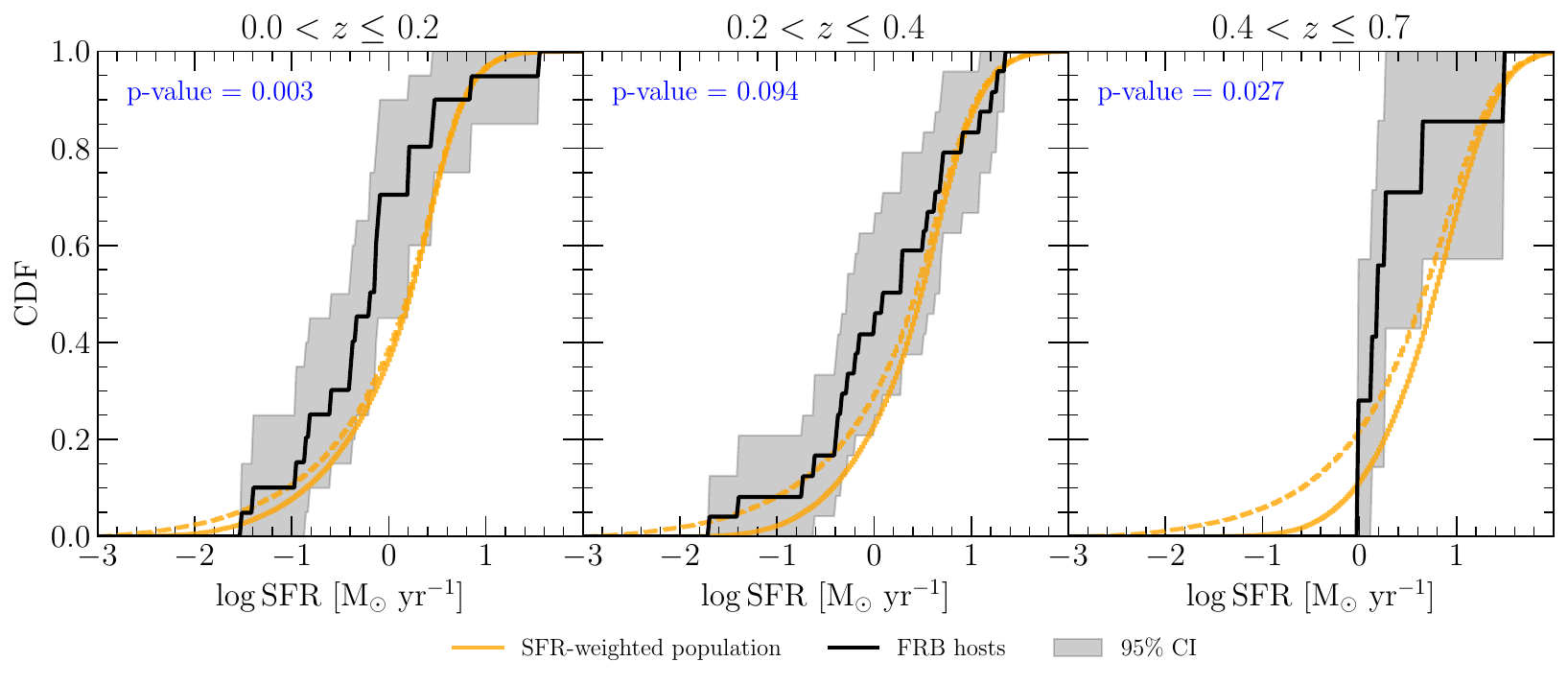}
    \caption{Comparison of CDF in $\mathrm{SFR}$ between mock generated, $\mathrm{SFR}$-weighted galaxy samples and FRB hosts. Black solid line corresponds to the FRB data (see Sect.~\ref{sec4.1}), while the gray-shaded area depicts the $95\%$ confidence interval obtained using bootstrapping. Orange dashed (solid) line shows the CDF of the $\mathrm{SFR}$-weighted population without (with) the magnitude cut.  The mock galaxy sample used here is the same as in Sect.~\ref{sec4.2}. p-values estimated using the KS-test are also displayed in each panel. Evidently, the two distributions are in agreement with each other across all redshift bins, constituting another line of evidence towards FRBs tracing SF, though the SFR-selected population exhibits systematically higher SFR values.}
    \label{App_fig1}
\end{figure*}

\section{Comparison of cumulative distributions in $\mathrm{SFR}$} \label{App.SFR_cdf}

In Sect.~\ref{sec4}, we focused on the CDF in stellar mass and provided lines of evidence in favor of FRBs tracking SF. Given that both FRB hosts and mock-generated populations have available information about $\mathrm{SFR}$, we also compare the CDFs in $\mathrm{SFR}$ and test whether we draw the same conclusion concerning the statement that FRBs trace $\mathrm{SFR}$. Nevertheless, measuring $\mathrm{SFR}$ through SED-fitting is challenging and can lead to remarkably different results, depending on the underlying assumptions of the model, such as the star formation history and the fitting tool implemented in the analysis. Conversely, the stellar mass inference is not very sensitive to the SED model that is used in the fit, as it scales with the total luminosity and therefore is considered more trustworthy than the $\mathrm{SFR}$.

For completeness, though, we compare the CDF of $\mathrm{SFR}$ between the observed FRB hosts and mock-generated, $\mathrm{SFR}$-weighted galaxy sample (identical to the one in Sect.~\ref{sec4.2}). The results are displayed in Fig.~\ref{App_fig1}. While the data favors systematically lower $\mathrm{SFR}$ values than the mock galaxy population, the two distributions are statistically consistent with each other given the small sample sizes. 
We quantified this with a KS-test; the null hypothesis that they are drawn from the same parent distribution cannot be rejected (within $\sim 3 \sigma)$. In particular, the p-value is equal to $0.003$, $0.094$, and $0.027$ in the range $0<z\leq 0.2$, $0.2<z\leq 0.4$, and $0.4<z\leq 0.7$, respectively.

\section{Comparison of CDFs obtained using different mass functions} \label{App.MF}

In this study, we made use of the stellar mass function offered by \cite{Leja2020}, which is described by a two-component Schechter function. For completeness, we repeat the calculations by implementing a single Schechter mass function (used in \citealt{Bochenek2021}; see also Eq.~\ref{eq.2}) to quantify the sensitivity of our conclusions to the choice of the stellar mass function in the mock galaxy's sample generator. We make this comparison with $\mathrm{SFR}$-weighted populations; using mass-weighted populations gives the same conclusions.

\begin{figure*}[h!]
    \centering
    \includegraphics[width=\textwidth]{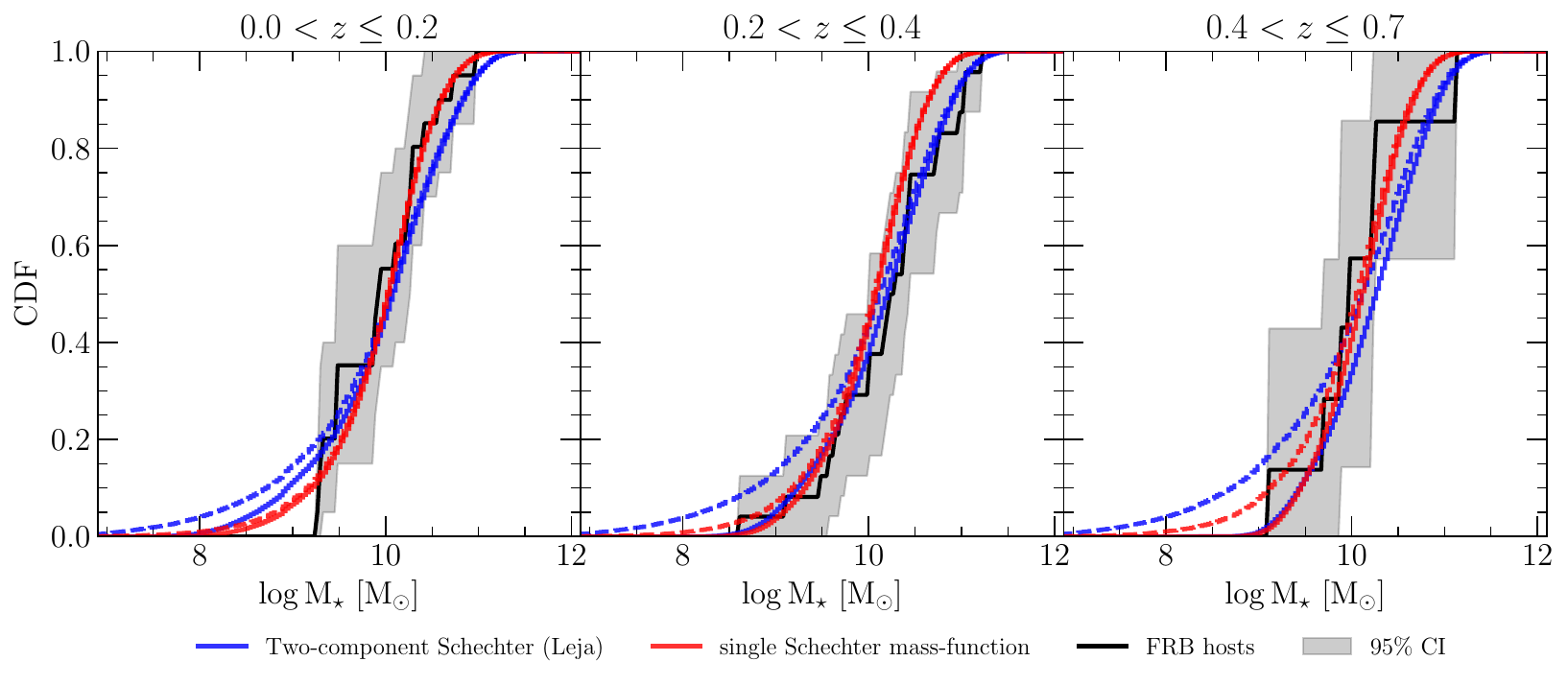}
    \caption{Comparison of CDF in mass between mock generated, $\mathrm{SFR}$-weighted populations obtained using different mass functions. Blue line shows the one based on the two-component Schechter mass function \citep{Leja2020} and is essentially identical to the blue line in Fig.~\ref{fig6}. Red line represents the population sampled using the single Schechter mass function (see Eq.~\ref{eq.2}). Dashed lines denote the CDF of the full population, while solid lines refer to the optically corrected ones. For visualization purposes we also include the FRB hosts' data considered in this paper (see caption in Fig.~\ref{fig6} for details).}
    \label{App_fig4}
\end{figure*}

The single Schechter function is flatter at lower masses, that is, it has a smaller number of low-mass galaxies, thus we expect the resulting CDF to be shifted to higher masses. Fig.~\ref{fig1} (left panel) gives a direct comparison between the two mass functions. Fig.~\ref{App_fig4} shows the resulting CDFs. The distribution corresponding to the single Schechter function rises at higher masses due to the suppression of the mass function at low masses, so the two populations yield significantly different CDFs. Nevertheless, once the samples are magnitude-limited (solid lines), the differences become less prominent owing to the fact that the low mass galaxies (the mass scale at which the differences in the two mass functions dominate) tend to be fainter and are more likely to drop out. While the CDFs are not identical, the two distributions lead to the result that FRBs are consistent with being SF tracers. Note that the predicted CDF based on the two-component Schechter function has a smaller slope. This is due to the fact that the $\mathrm{SFR}$ and/or mass-weighted mass function is flatter, yielding both low mass and high mass galaxies in the sample than the single Schechter one.

\bibliography{references}{}
\bibliographystyle{aasjournal}

\end{document}